\documentclass[aip,reprint,jcp,amsmath,amssymb,showpacs,superscriptaddress,nofootinbib]{revtex4-2}
\usepackage{bm}
\usepackage{graphicx}
\usepackage{color}
\usepackage{appendix}
\usepackage{hyperref}
\usepackage{microtype}
\definecolor{deepblue}{rgb}{0,0,0.5}
\definecolor{deepred}{rgb}{0.6,0,0}
\hypersetup{
colorlinks,%
citecolor=deepblue,%
filecolor=black,%
linkcolor=deepblue,%
urlcolor=deepblue
}
\newcommand{\red}[1]{#1}

\newcommand{\mat}[1]{{\bf {#1}}}

\newcommand{\odif}[2]{\frac{d #1}{d #2}}
\newcommand{\oodif}[2]{\frac{d^2 #1}{d {#2}^2}}
\newcommand{\pdif}[2]{\frac{\partial #1}{\partial #2}}
\newcommand{\ppdif}[2]{\frac{\partial^2 #1}{\partial {#2}^2}}
\newcommand{\ave}[1]{\langle {#1} \rangle}
\newcommand{\Ave}[1]{\Bigl\langle {#1} \Bigr\rangle}

\graphicspath{{plots/}:{./}}

\begin{document}

\title{Revisiting the single-saddle model for the $\beta$-relaxation of supercooled liquids}

\author{Daniele Coslovich}
\email{dcoslovich@units.it}
\affiliation{Dipartimento di Fisica, Universit\`a di Trieste, Strada Costiera 11, 34151, Trieste, Italy}

\author{Atsushi Ikeda}
\email{atsushi.ikeda@phys.c.u-tokyo.ac.jp}
\thanks{The following article has been accepted by The Journal of Chemical Physics. After it is published, it will be found at \url{https://aip.scitation.org/journal/jcp}.}
\affiliation{Graduate school of arts and science, University of Tokyo, Komaba, Tokyo 153-8902, Japan}
\affiliation{Research Center for Complex Systems Biology, Universal Biology Institute, University of Tokyo, Komaba, Tokyo 153-8902, Japan}

\date{\today}

\begin{abstract}
The dynamics of glass-forming liquids display several outstanding features, such as two-step relaxation and dynamic heterogeneities, which are difficult to predict quantitatively from first principles.  
In this work, we revisit a simple theoretical model of the $\beta$-relaxation, \textit{i.e.}, the first step of the relaxation dynamics.
The model, first introduced by Cavagna \textit{et al.}, describes the dynamics of the system in the neighborhood of a saddle point of the potential energy surface.
We extend the model to account for density-density correlation functions and for the 4-point dynamic susceptibility.
We obtain analytical results for a simple schematic model, making contact with related results for $p$-spin models and with the predictions of inhomogeneous mode-coupling theory.
Building on recent computational advances, we also explicitly compare the model predictions against overdamped Langevin dynamics simulations of a glass-forming liquid close to the mode-coupling crossover.
The agreement is quantitative at the level of single-particle dynamic properties \red{only} up to the early $\beta$-regime.
Due to its inherent harmonic approximation, however, the model is unable to predict the dynamics on the time scale relevant for structural relaxation.
Nonetheless, our analysis suggests that the agreement with the simulations may be largely improved if the modes' spatial localization is properly taken into account.
\end{abstract}

\maketitle

%%%%%%%%%%%%%%%%%%%%%%%%%%%%%%%%%%%%%%%%%%%%%%%%%%%%%%%%%%%%%%%%%%%%
\section{Introduction}
%%%%%%%%%%%%%%%%%%%%%%%%%%%%%%%%%%%%%%%%%%%%%%%%%%%%%%%%%%%%%%%%%%%%

Predicting the dynamical properties of supercooled liquids from first principles is possibly one of the hardest challenges in theoretical condensed matter physics~\cite{cavagna_supercooled_2009,berthier_theoretical_2011}. 
In this context, ``first principles'' refers to a theory that starts from the exact microscopic equations of motion of the system of interest and contains no adjustable parameters. 
Mode-coupling theory (MCT)~\cite{gotze_complex_2009} is probably the most well-known, first-principles theory of the dynamics of supercooled liquids. 
It accounts for several nontrivial features, such as the presence of two-step relaxation or the shape of the non-ergodicity parameters, but also predicts a spurious divergence of the structural relaxation time at a temperature $T_\textrm{MCT}$ at which the liquid is still fully ergodic. 
A common interpretation is that the sharp transition predicted at $T_\textrm{MCT}$ is smeared by thermal activation, which is not accounted for by the theory and turns the transition into a crossover. 
A systematic way to improve MCT is to take into account higher order correlations and several attempts along this line have been made~\cite{Szamel_2003,Mayer2006,janssen_microscopic_2015,luo_tagged-particle_2021,ciarella_multi-component_2021}.
Recent advances have also improved our understanding of the slow dynamics of liquids in higher dimensions~\cite{Baity-Jesi_Reichman_2019,Berthier_Charbonneau_Kundu_2020} and an exact solution for the dynamics of hard hyper-spheres in the infinite dimensional limit has been found~\cite{Maimbourg_Kurchan_Zamponi_2016}.

A central concept for the theoretical description of supercooled liquids is the so-called potential energy surface (PES)~\cite{Stillinger1982,wales2003energy,Sciortino_2005,Heuer2008}. 
The PES is defined by the total potential energy $\mathcal{V}$ as a function of the configurational state of the system. Configuration space can then be partitioned into basins of attractions associated to the local minima of $\mathcal{V}$. Through a statistical description of the properties of such basins, it is possible to develop a quantitative formalism, which successfully accounts for the thermodynamic properties of supercooled liquids~\cite{wales2003energy,Sciortino_2005}. 
Predicting the dynamics from the statistical properties of the PES is, however, a much more challenging task~\cite{Heuer2008,Doye_Saddle_2002}. 
At the end of the 1990's, a series of numerical studies~\cite{Stratt_1995,Bembenek_Laird_1995,Keyes_Vijayadamodar_Zurcher_1997,Ribeiro_Madden_1997,Kramer_Buchner_Dorfmuller_1998} led to a first principles description of the liquid dynamics in terms of so-called instantaneous normal modes, obtained by diagonalizing the Hessian matrix of the potential energy at equilibrium configurations. However, the extension of these ideas to supercooled liquids encountered some difficulties \cite{Gezelter_Rabani_Berne_1997} and the approach remained largely phenomenological~\cite{Donati_Sciortino_Tartaglia_2000}, see Refs.~\onlinecite{Clapa_Kottos_Starr_2012,Zhang_Douglas_Starr_2019,Kriuchevksyi_Sirk_Zaccone_2021} for recent developments.

A further attempt to develop a first principles, PES-based description of the dynamics is due to Cavagna \textit{et al.}~\cite{cavagna_single_2003}, who introduced a model of the so-called $\beta$-regime, \textit{i.e.}, the first step of the relaxation. Contrary to the instantaneous normal mode approach, the single-saddle model (SSM) of Cavagna \textit{et al.} focused on stationary points of the PES with a finite number $n_u$ of unstable modes. The key hypothesis was that above the MCT crossover temperature, the motion of the system in configuration space mostly follows the unstable directions of nearby saddles. A local harmonic expansion around those points should therefore provide information on the mean square displacement of the particles at short times.
The predictions of the SSM were, however, never tested against computer simulation results.
Moreover, while saddle-based approaches were successful in describing the dynamical transition in mean-field $p$-spin models \cite{cavagna_supercooled_2009}, they faced some technical and conceptual difficulties when applied to finite-dimensional structural glasses \cite{grigera_geometric_2002,wales_dynamics_2001,Wales_Doye_2003,Grigera_2006}.
Since most of these issues have recently been solved \cite{coslovich_localization_2019,Shimada_Coslovich_Mizuno_Ikeda_2021}, we think that the times are ripe to revisit in greater detail saddle-based approaches to the dynamics of supercooled liquids.

In this work, we provide a systematic assessment of the predictions of the SSM and compare them against results of computer dynamics simulations of a realistic model glass.
We work out in full detail the SSM expressions for the density-density correlation functions and for the 4-point dynamic susceptibility. 
A simple schematic version of the SSM reveals a connection with the dynamic scaling predicted by the so-called inhomogeneous MCT~\cite{Biroli_Bouchaud_Miyazaki_Reichman_2006}. 
The comparison of the theoretical predictions with the Langevin dynamics simulations shows that the SSM provides an accurate description of spatio-temporal correlations in the early $\beta$-regime, \text{i.e.}, the approach to the plateau, in particular for the single-particle dynamics.
At longer times, however, the theoretical description is not satisfactory, due to the harmonic approximation inherent in the model.
We finally discuss possible ways to improve the agreement between the model and the simulation data.

%%%%%%%%%%%%%%%%%%%%%%%%%%%%%%%%%%%%%%%%%%%%%%%%%%%%%%%%%%%%%%%%%%%%
\section{The single saddle model}\label{sec:ssm}
%%%%%%%%%%%%%%%%%%%%%%%%%%%%%%%%%%%%%%%%%%%%%%%%%%%%%%%%%%%%%%%%%%%%
We consider $N$ interacting Brownian particles in a $d$-dimensional cell with periodic boundary conditions. 
Let us first summarize our notation. 
We use right arrow vectors to express vectors in the $d$-dimensional space and boldface vectors to express vectors in the $dN$-dimensional configuration space: $\vec{r}_i(t)$ denotes the position of particle $i$ at time $t$, while $\bm{r}(t) = (\vec{r}_1(t),...,\vec{r}_N(t))$ denotes the position of the system in the configuration space. 
We use $i, j...$ for the particle index and $a, b... $ for the configuration space index, \textit{e.g.}, $a = 1x$ means the coordinate $x$ of particle 1. 
We also use the notation $a \in i$ to express the subset of the configuration space indices corresponding to particle $i$.

The microscopic time evolution of the system is given by the overdamped Langevin equation
\begin{eqnarray}
\gamma \odif{\bm{r}}{t} = - \pdif{\mathcal{V}}{\bm{r}} + \bm{\eta}, \label{eom}
\end{eqnarray}
where $\mathcal{V}(\bm{r})$ is the potential energy of the system, $\gamma$ is the damping coefficient, and $\bm{\eta}(t)$ is the random Gaussian noise at time $t$, \textit{i.e.}, $\ave{\eta_a(t)} = 0$ and $\ave{\eta_a(t)\eta_b(t')} = 2 \gamma k_B T \delta_{ab} \delta(t-t')$ where $k_B$ is the Boltzmann constant and $T$ is the temperature. 
The average over realizations of the noise, for a given initial configuration $\bm{r}_0$, is denoted by $\ave{\cdots}$. 
We set $\gamma = 1$ and $k_B = 1$ to fix the units of time and temperature. 

We consider the situation in which the initial configuration is a stationary point of the energy landscape $\mathcal{V}(\bm{r})$, either a local minimum or a saddle. 
We then focus on the time evolution of the displacements $\bm{x}(t) \equiv \bm{r}(t) - \bm{r}_0$. 
We expand the potential energy as $\mathcal{V}(\bm{r}) = \frac{1}{2} \bm{x} \cdot \mat{M} \cdot \bm{x} + \mathcal{O}(x^3)$, where $\mat{M}$ is the dynamical matrix for the initial configuration
\begin{eqnarray}
M_{ab} = \left. \frac{\partial^2 \mathcal{V}}{\partial r_a \partial r_b} \right|_{\bm{r} = \bm{r}_0}. 
\end{eqnarray}
Inserting this expansion into the Langevin equation and omitting higher order terms, we obtain the harmonic equations of motion 
\begin{eqnarray}
\odif{x_a}{t} = - \sum_b M_{ab} x_b + \eta_a. \label{Leq}
\end{eqnarray}
The dynamics described by these equations of motion defines the SSM. 
The central quantity in this model is the dynamical matrix $\mat{M}$. 
Let $\lambda_\alpha$ and $\bm{e}_\alpha$ denote the $\alpha$-th eigenvalue and eigenvector, respectively. 
We use $\alpha, \beta...$ for the index of the eigenmodes $\alpha = 1,...,dN$. 
Note that eigenvectors are orthonormalized: $\bm{e}_{\alpha} \cdot \bm{e}_{\beta} = \delta_{\alpha \beta}$. 

The SSM was introduced by Cavagna \textit{et al.}~\cite{cavagna_single_2003} to predict the mean square displacement (MSD) in a supercooled liquid. 
Here, we extend this earlier work to calculate the intermediate scattering functions as well as the four-point dynamic susceptibility. 
To this end, we consider the corresponding Fokker-Planck equation 
\begin{eqnarray}
\pdif{}{t} P(\bm{x},t) = \sum_{ab} \pdif{}{x_a} \left[ M_{ab} x_b  P(\bm{x},t) \right] + T \sum_a \ppdif{}{x_a} P(\bm{x},t), \nonumber \\ 
\end{eqnarray}
where is $P(\bm{x},t)$ the probability density for the displacement $\bm{x}$ at time $t$. 
We are interested in the solution of this equation with the initial condition $P(\bm{x},0) = \delta(\bm{x})$. 
This is achieved by Fourier transformation~\cite{zwanzig2001nonequilibrium}; the solution is
\begin{eqnarray}
P(\bm{x},t) = \frac{1}{\sqrt{\det( 2\pi \mat{S}(t))}} \mathrm{e}^{- \frac{1}{2} \bm{x} \cdot \mat{S}^{-1}(t) \cdot \bm{x}}, \label{fpsol}
\end{eqnarray}
where 
\begin{eqnarray}\label{eqn:S_and_K}
\mat{S}(t) = T \sum_{\alpha} K(\lambda_{\alpha},t) \bm{e}_\alpha \bm{e}_\alpha
\end{eqnarray}
with $K(\lambda_{\alpha},t) = \frac{1 - \mathrm{e}^{-2 \lambda_\alpha t}}{\lambda_\alpha}$. 
Note that $\mat{S}(t)$ is a $dN \times dN$ symmetric matrix, and it is positive-definite because $K(\lambda_{\alpha},t) > 0$ for any real $\lambda_{\alpha}$ at $t >0$. 
We use this solution to calculate several correlation functions of interest.

Standard correlation functions to probe the dynamics of supercooled liquids are the MSD and the self and collective intermediate scattering functions, which are defined by 
\begin{eqnarray}
&& \hat{R}(t) = \frac{1}{N} \sum_i |\vec{x}_i(t)|^2, \\
&& \hat{F}_s(\vec{k},t) = \frac{1}{N} \sum_i \cos (\vec{k} \cdot \vec{x}_i(t)), \\
&& \hat{F}(\vec{k},t) = \frac{1}{N} \sum_{ij} e^{i \vec{k} \cdot (\vec{r}_i(t) - \vec{r}_j(0))}, 
\end{eqnarray}
\red{for a single trajectory starting from a given initial configuration.}
In the SSM, the average over noise can be expressed using eigenvalues and eigenvectors. 
The MSD can be calculated as: 
\begin{eqnarray}\label{eqn:msd}
\ave{\hat{R}(t)} = \frac{1}{N} \sum_i \sum_{a \in i} \int d\bm{x} x_a^2  P(\bm{x},t)
%= \sum_{a} S_{aa}(t) 
= \frac{T}{N} \sum_{\alpha} K(\lambda_{\alpha},t). \nonumber \\
\end{eqnarray}
This expression was already obtained in Ref.~\onlinecite{cavagna_single_2003}.
Starting from the Fourier transform of Eq.~\eqref{fpsol}, we can also obtain the expressions for the intermediate scattering functions (see Appendix~\ref{sec:appendix1}): 
\begin{eqnarray}
&& \ave{\hat{F}_s(\vec{k},t)} = \frac{1}{N} \sum_i \mathrm{e}^{- \frac{T}{2} \sum_{\alpha} K(\lambda_{\alpha},t) (\vec{k} \cdot \vec{e}_{\alpha,i})^2}, \\
&& \ave{\hat{F}(\vec{k},t)} = \frac{1}{N} \sum_{ij} \mathrm{e}^{i \vec{k} \cdot (\vec{r}_{i,0} - \vec{r}_{j,0})} 
\mathrm{e}^{- \frac{T}{2} \sum_{\alpha} K(\lambda_{\alpha},t) (\vec{k} \cdot \vec{e}_{\alpha,i})^2},\nonumber\\
\label{fktresult}
\end{eqnarray}
where $\vec{e}_{\alpha,i}$ is the $i$-th particle contribution to the $\alpha$-th eigenvector $\bm{e}_{\alpha}$. 
Note that the MSD can be calculated using only the eigenvalues, while the intermediate scattering functions depend explicitly on the eigenvectors. 

Another important quantity to characterize the supercooled dynamics is the four-point dynamic susceptibility. 
In particular, we consider two different forms of it: 
\begin{eqnarray}
&& \hat{\chi}_{R,{\rm iso}}(t) = N \left[ \hat{R}(t)^2 - \ave{\hat{R}(t)} \right]^2, \label{chirdef} \\
&& \hat{\chi}_{4,{\rm iso}}(\vec{k},t) = N \left[ \hat{F}_s(\vec{k},t)^2 - \ave{\hat{F}_s(\vec{k},t)} \right]^2. 
\end{eqnarray}
The fluctuations of $\hat{F}_s(\vec{k},t)$ have been frequently investigated in computational studies of supercooled liquids~\cite{berthier_theoretical_2011}, while those of $\hat{R}(t)$ have been used to characterize the anomalous vibrations near the jamming transition~\cite{Ikeda2013}. 
\red{We emphasize that the susceptibilities $\langle \hat{\chi}_{R,{\rm iso}}(\vec{k},t)\rangle$ and $\langle \hat{\chi}_{4,{\rm iso}}(\vec{k},t)\rangle$ are computed in the so-called isoconfigurational ensemble~\cite{Widmer-Cooper_Harrowell_Fynewever_2004}, in which only the fluctuations of trajectories starting from the same configuration are taken into account, while the full dynamic susceptibility has an additional contribution coming from sample-to-sample fluctuations~\cite{Berthier_Jack_2007,Franz2011a}}.
The SSM expression for $\ave{\hat{\chi}_{R,{\rm iso}}(t)}$ can be obtained in a similar way as $\ave{\hat{R}(t)}$: 
\begin{eqnarray}
\ave{\hat{\chi}_{R,{\rm iso}}(t)} &=& \frac{1}{N} \sum_{ab} \int d\bm{x} x_a^2 x_b^2 P(\bm{x},t) - N R(t)^2 \nonumber \\
&=& \frac{2}{N} \sum_{ab} S_{ab}(t)^2
= \frac{2T^2}{N} \sum_{\alpha} K(\lambda_{\alpha},t)^2, \label{chir}
\end{eqnarray}
where the identity $\sum_{a} e_{\alpha,a} e_{\beta,a} = \delta_{\alpha \beta}$ was used in the final line. 
As shown in Appendix~\ref{sec:appendix1}, $\ave{\hat{\chi}_{4,{\rm iso}}(\vec{k},t)}$ can be calculated in a similar way as $\ave{\hat{F}_s(\vec{k},t)}$, and we obtain 
\begin{eqnarray}
\ave{\hat{\chi}_{4,{\rm iso}}(\vec{k},t)} &=& \frac{1}{N} \sum_{ij} \left[ 
\frac{1}{2} \mathrm{e}^{- \frac{T}{2} \sum_{\alpha} K(\lambda_{\alpha},t) (\vec{k} \cdot (\vec{e}_{\alpha,i} + \vec{e}_{\alpha,j}))^2} \right. \nonumber \\ 
&& + \frac{1}{2} \mathrm{e}^{- \frac{T}{2} \sum_{\alpha} K(\lambda_{\alpha},t) (\vec{k} \cdot (\vec{e}_{\alpha,i} - \vec{e}_{\alpha,j}))^2} \nonumber \\
&& \left. - \mathrm{e}^{- \frac{T}{2} \sum_{\alpha} K(\lambda_{\alpha},t) ((\vec{k} \cdot  \vec{e}_{\alpha,i})^2 + (\vec{k} \cdot \vec{e}_{\alpha,j})^2)} \right]. \label{chi4result}
\end{eqnarray}
\red{We note that the averaged correlation functions and susceptibilities calculated in this section still depend on the initial configuration $\bm{r}_0$. 
We denote the average over initial configurations within some ensemble by $\ave{\cdots}_c$.
In the following, we will remove the hat symbol only after the averages over both noise and initial configurations are taken, \textit{e.g.}, $A = \ave{\ave{\hat{A}}}_c$.}

%%%%%%%%%%%%%%%%%%%%%%%%%%%%%%%%%%%%%%%%%%%%%%%%%%%%%%%%%%%%%%%%%%%%
\section{Schematic model}
%%%%%%%%%%%%%%%%%%%%%%%%%%%%%%%%%%%%%%%%%%%%%%%%%%%%%%%%%%%%%%%%%%%%

\subsection{Setting}

The dynamics of the SSM depends on \red{the initial configuration $\bm{r}_0$, in particular through} the dynamical matrix $\mat{M}$. 
In this section, we consider a schematic model of an ensemble of $\mat{M}$, which was introduced in Ref.~\onlinecite{cavagna_single_2003}. 
We first set $d=1$ as it becomes clear that the spatial dimension plays no role in this simple model. 
Accordingly, $\mat{M}$ is $N \times N$ matrix and the particle indexes $i,j...$ are equivalent with the configuration space indexes $a,b...$. 
We assume that the eigenvalues are distributed according to the semi-circle law:    
\begin{eqnarray}
\rho (\lambda) = \begin{cases}
\frac{2}{\pi} \sqrt{1 - (\lambda + \epsilon -1)^2} & (-\epsilon \leq \lambda \leq 2- \epsilon) \\ 
0 & (\mbox{otherwise}).
\end{cases}
\label{circle}
\end{eqnarray}
The minimum eigenvalue is $- \epsilon$. 
When $\epsilon > 0$, $\rho (\lambda)$ has a negative support, which corresponds to saddles, while when $\epsilon \leq 0$, $\rho (\lambda)$ has only positive support, which corresponds to local minima. 
This assumption holds if $N$ is asymptotically large and $\mat{M}$ is a symmetric random matrix drawn from the Gaussian ensemble~\cite{mehta2004random} plus the diagonal matrix $(1 - \epsilon) \mat{1}$. 
This schematic model is frequently encountered in mean field disordered systems, \textit{e.g.}, the statistical properties of the saddles of the $p$-spin spherical model follow these assumptions, where $\epsilon$ plays a role of the deviation of the temperature from the dynamical transition temperature: $\epsilon \propto T/T_c - 1$~\cite{Cavagna1998}. 

To calculate the wave-vector dependent quantities, we further assume that the components of the eigenvectors are Gaussian random variables
\begin{eqnarray}
f(e_i) = \sqrt{\frac{N}{2\pi}}e^{- N e_i^2/2}. \label{uniform}
\end{eqnarray}
In the limit $N \to \infty$, the eigenvector $\bm{e}$ is distributed uniformly on the $N$-dimensional unit sphere, as in the case of the Maxwell distribution of velocity of hard spheres in the microcanonical ensemble. 

We calculate the correlation functions within the schematic model in the thermodynamic limit $N \to \infty$.
When using the expressions derived in the previous section, we set $T=1$, because the main effect of temperature is encoded in the eigenvalue distribution through $\epsilon$. 
\red{Note that the schematic model yields correlation functions averaged over both the realizations of the noise and dynamical matrices $\mat{M}$, which corresponds to the double average $\ave{\ave{\cdots}}_c$ in our notation. } 

\subsection{Mean square displacement and dynamic susceptibility}

\begin{figure}
    \centering
    \includegraphics[scale=0.98]{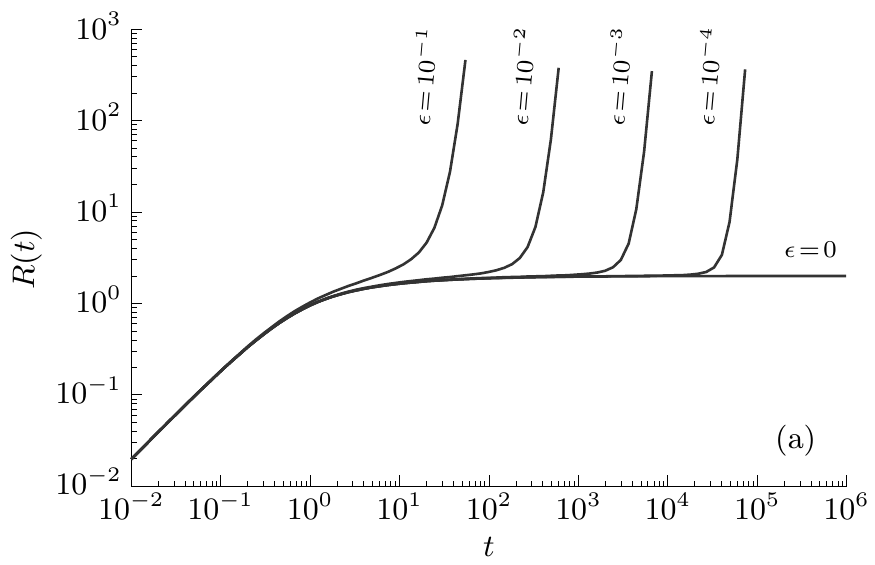}
    \includegraphics[scale=0.98]{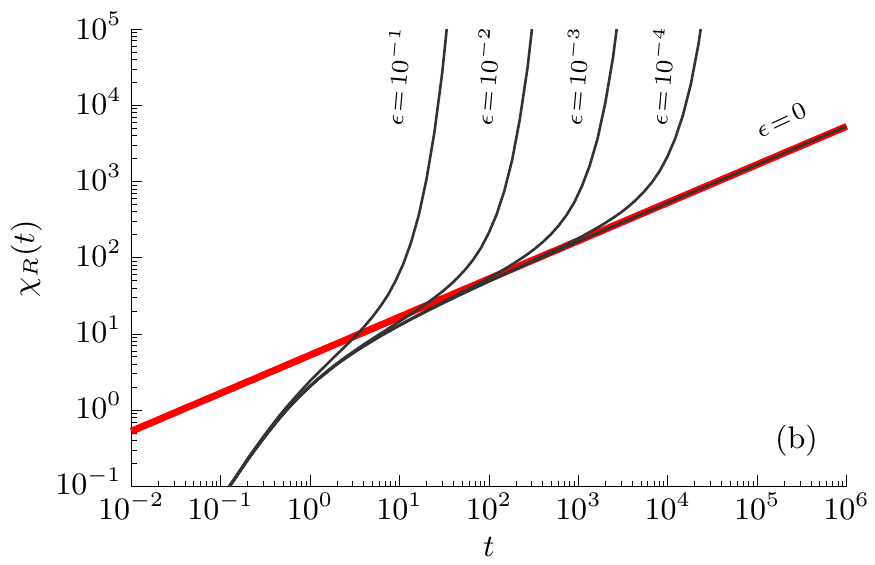}
    \caption{
(a) Mean-square displacement $R(t)$ and (b) four-point dynamic susceptibility of displacements $\chi_R(t)$ in the schematic model. The numerical results for $\epsilon = 0, 10^{-4}, 10^{-3}, 10^{-2},10^{-1}$ are plotted.
The thick red line in (b) indicates the asymptotic formula Eq.~\eqref{chi4beta} for the $\beta$-regime, which is proportional to $t^{1/2}$. }
    \label{fig:modelmsd}
\end{figure}

$R(t)$ and $\chi_{R,{\rm iso}}(t)$ can be calculated without using the eigenvectors. 
By inserting the spectra Eq.~\eqref{circle}, we obtain 
\begin{eqnarray}
&& R(t) 
= \frac{2}{\pi} \! \int_{-\epsilon}^{2-\epsilon} \!\!\!\! d\lambda \left( \frac{1 - \mathrm{e}^{-2 \lambda t}}{\lambda} \right) \! \sqrt{1 - (\lambda + \epsilon -1)^2} \label{msdrm} \\ 
&& \chi_{R,{\rm iso}}(t) 
= \frac{4}{\pi}\! \int_{-\epsilon}^{2-\epsilon} \!\!\!\! d\lambda \left( \frac{1 - \mathrm{e}^{-2 \lambda t}}{\lambda} \right)^2 \!\!\!\! \sqrt{1 - (\lambda + \epsilon -1)^2}. \nonumber \\ \label{chirrm}
\end{eqnarray}
We numerically computed these integrals using the trapezoidal rule. 
The results are shown in Fig.~\ref{fig:modelmsd}. 
At short times, the MSD shows a diffusive behavior $R(t) = 2t$, which correponds to the non-interacting regime and can be reproduced by setting $\mat{M} = 0$ in the equation of motion Eq.~\eqref{Leq}. 
Then, $R(t)$ approaches a plateau, corresponding to the $\beta$-relaxation regime of MCT, from which it exits on a time scale that diverges as $\epsilon$ decreases.
The long-time limit of the MSD, $R_{\infty}$, can be calculated by setting $\epsilon = 0$ and $t \to \infty$ in Eq.~\eqref{msdrm}: 
\begin{eqnarray}
R_{\infty}
= \frac{2}{\pi} \int_{0}^{2} d\lambda \sqrt{\frac{2 - \lambda}{\lambda}}
= 2.
\end{eqnarray}
The behavior in the $\beta$-regime is qualitatively similar to the one found in supercooled liquids. 
However, $R(t)$ grows exponentially at long times, because of the factor $\mathrm{e}^{-2\lambda t}$ for the negative $\lambda$: this unphysical behavior is obviously due to a breakdown of the local harmonic approximation. 
Therefore, the validity of the SSM is limited to the $\beta$-regime.

The four-point dynamic susceptibility of displacements behaves as $\chi_{R,{\rm iso}}(t) = 8t^2$ in the short-time, non-interacting regime.
In the early $\beta$-relaxation regime, $\chi_R(t)$ shows a power-law growth $\chi_{R,{\rm iso}}(t) \propto t^{1/2}$. 
Finally, it grows exponentially in the $\alpha$ relaxation regime. 
Interestingly, $\chi_{R,{\rm iso}}(t)$ does not stop growing even at $\epsilon =0$. 

%\subsection{$F_s(k,t)$, $F(k,t)$, and $\chi_4(k,t)$}
\subsection{Intermediate scattering functions and dynamic susceptibility}

We now focus on the wave-vector dependent quantities $F_s(k,t)$, $F(k,t)$, and $\chi_{4,{\rm iso}}(k,t)$. 
In the schematic model, we can calculate $F_s(k,t)$ in the following way 
\begin{eqnarray}
F_s(k,t) 
&=& \left[ \int d\lambda \rho(\lambda) \left(1 + \frac{k^2 K(\lambda,t)}{N} \right)^{-1/2} \right]^N \nonumber \\
&=& \left[ 1 - \frac{k^2 R(t)}{2N} + O(N^{-2}) \right]^N
\!\!\!\!\! = \mathrm{e}^{- k^2 R(t)/2}. 
\end{eqnarray}
In the final line, we expanded the square root and took the $N \to \infty$ limit. 
We thus recover a simple relation between $R(t)$ and $F_s(k,t)$, known as the Gaussian approximation for $F_s(k,t)$ in the context of finite dimensional liquids~\cite{hansen_theory_2006}. 

We can calculate $F(k,t)$ in a similar manner and we obtain
\begin{eqnarray}
F(k,t) = S(k) \mathrm{e}^{- k^2 R(t)/2},
\end{eqnarray}
where $S(k) = \frac{1}{N} \sum_{ij} \mathrm{e}^{i \vec{k} \cdot (\vec{r}_{i,0} - \vec{r}_{j,0})}$ is the static structure factor.
Therefore in the schematic model, the self and collective intermediate scattering function exactly follow the relation
\begin{equation}\label{eqn:vineyard}
F(k,t) = S(k) F_s(k,t),
\end{equation}
which is the well-known Vineyard approximation~\cite{Vineyard_1958}. 
Note that this relation holds in the schematic model but not generally in the SSM, because the SSM allows for correlations between configurations $\bm{r}$ and eigenvectors $\bm{e}$. 
We will discuss this point further in Sec.~\ref{sec:supercooled} 

We finally calculate the four-point dynamic susceptibility. 
The calculation goes in a similar way as $F_s(k,t)$ but is a bit lengthy, see Appendix~\ref{sec:appendix2}. The result is
\begin{eqnarray}
\chi_{4,{\rm iso}}(k,t) = \frac{1}{2} (1 - \mathrm{e}^{- k^2 R(t)})^2 + \frac{1}{4} k^4 \chi_{R,{\rm iso}}(t) \mathrm{e}^{- k^2 R(t)}.  \nonumber \\ \label{chi4rdm}
\end{eqnarray}
Therefore, in the schematic model, $F_s(k,t)$ and $\chi_{4,{\rm iso}}(k,t)$ can be written in terms of $R(t)$ and $\chi_{R,{\rm iso}}(t)$ only. 

\begin{figure}
    \centering
    \includegraphics[scale=0.98]{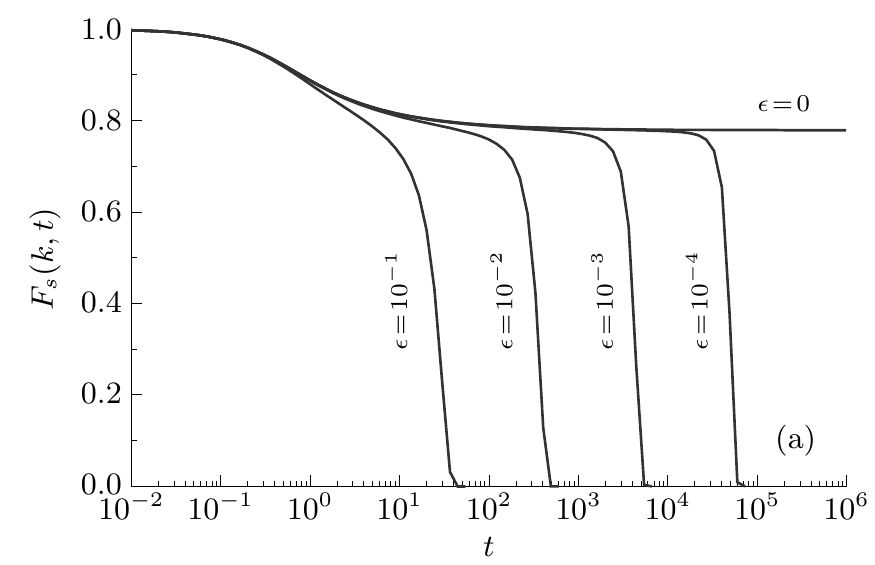}
    \includegraphics[scale=0.98]{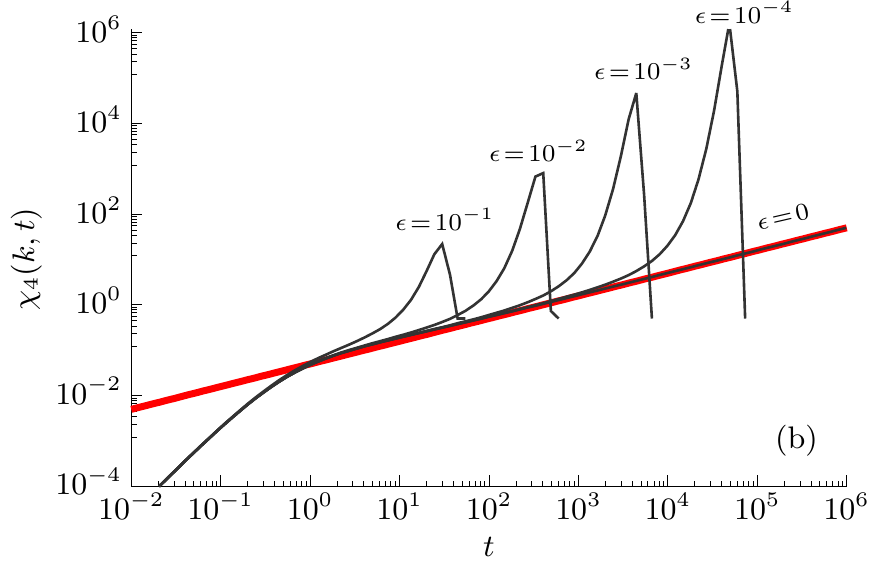}
    \caption{
(a) Self-part of the intermediate scattering function $F_s(k,t)(t)$ and (b) four-point dynamic susceptibility $\chi_4(k,t)$ in the schematic model. 
The numerical results for $k=1/2$ and $\epsilon = 0, 10^{-4}, 10^{-3}, 10^{-2},10^{-1}$ are plotted. 
The thick red line in (b) indicates the asymptotic formula Eq.~\eqref{chi4beta} for the $\beta$-regime, which is proportional to $t^{1/2}$. 
}
    \label{fig:modelfskt}
\end{figure}

The numerical results of $F_s(k,t)$ and $\chi_{4,{\rm iso}}(k,t)$ for $k=1/2$ are shown in Fig.~\ref{fig:modelfskt}. 
$F_s(k,t)$ qualitatively reproduces the canonical, two-step relaxation behavior of supercooled liquids. 
As $\epsilon$ approaches 0, the relaxation time scale diverges. 
The plateau height, also known as non-ergodicity parameter, is $F_{s,\infty}(k) = e^{-k^2}$ in this model since the long time limit of the MSD is $R_{\infty} = 2$. 
However, $F_s(k,t)$ shows a compressed exponential relaxation in the $\alpha$-relaxation regime, which is again due to the missing diffusive behavior of $R(t)$ discussed in the previous subsection: since $R(t)$ diverges exponentially, $F_s(k,t)$ decreases in a double exponential fashion.

The dynamic susceptibility $\chi_{4,{\rm iso}}(k,t)$ in the schematic model is also qualitatively similar to the one of supercooled liquids. 
It increases even in the $\beta$-relaxation regime, exhibits a peak in the $\alpha$ relaxation regime, and finally converges to 1/2, as observed in computer simulations of supercooled liquids~\cite{berthier_theoretical_2011}.
The long time limit $1/2$ originates from the self part; the distinct part goes to zero due to the factor $\mathrm{e}^{-k^2R(t)}$.
The growth observed in the $\beta$-relaxation regime follows $\chi_{4,{\rm iso}}(k,t) \propto t^{1/2}$, which is the same behavior of $\chi_{R,{\rm iso}}(t)$, as expected from Eq.~\eqref{chi4rdm}. 
This behavior will be further discussed in the next subsection.  

\subsection{Asymptotic analysis and discussion}

The power-law growth of $\chi_{4,{\rm iso}}(k,t)$ in the $\beta$-relaxation regime is reminiscent of computer simulation results~\cite{berthier_theoretical_2011} and of the prediction by the inhomogeneous MCT~\cite{Biroli2006}.
To rationalize this behavior, we perform an asymptotic analysis of $\chi_{4,{\rm iso}}(k,t)$.

\begin{figure}
    \centering
    \includegraphics[scale=0.98]{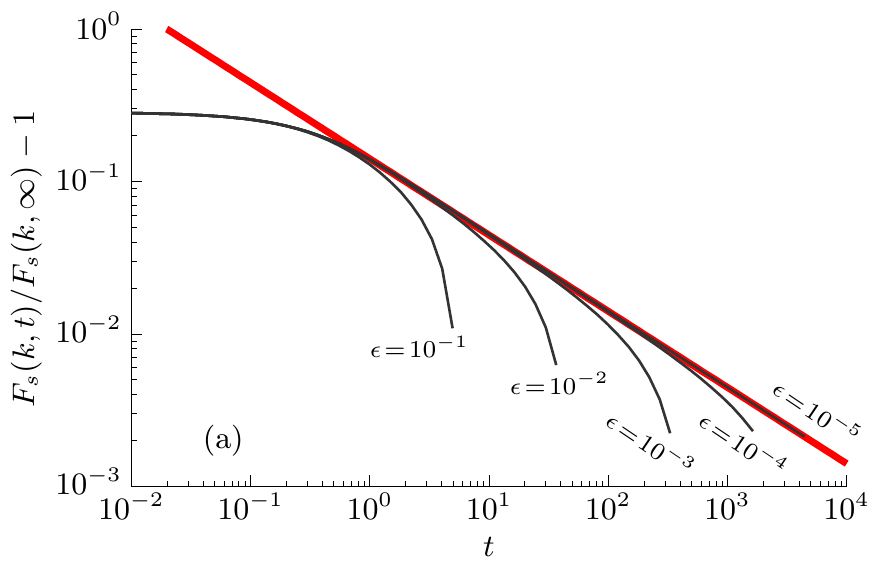}
    \includegraphics[scale=0.98]{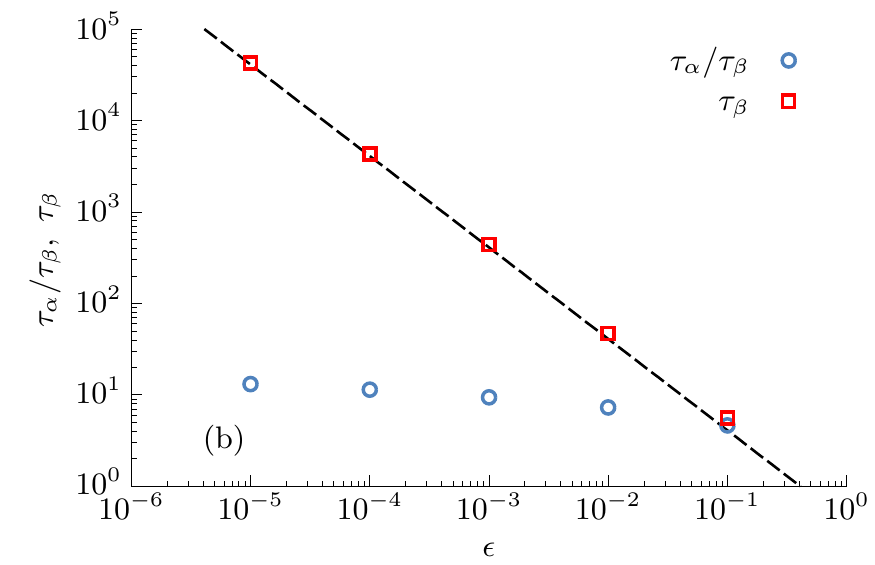}
    \caption{Test of the $\beta$ scaling for $F_s(k,t)$ for $k=1/2$ using the data in Fig. 2.
(a) The power-law behavior of $F_s(k,t)$ in the $\beta$-relaxation regime. 
The thick red line indicates the asymptotic formula Eq.~\eqref{fsktbeta} for the $\beta$-regime, which is proportional to $t^{-1/2}$. 
(b) Divergence of the relaxation time. $\tau_{\beta}$ and $\tau_{\alpha}/\tau_{\beta}$ are plotted against $\epsilon$. 
The dashed line indicates $0.41/\epsilon$. }
    \label{fig:asymptotic}
\end{figure}

We start from the asymptotic analysis of $R(t)$, which was already discussed in Ref.~\onlinecite{cavagna_single_2003}. 
Since we focus on the $t \gg 1$ region, it is sufficient to focus on the $\lambda \ll 1$ portion of the spectrum, which we approximate as $\rho (\lambda) = \frac{2}{\pi} \sqrt{2(\lambda + \epsilon)}$. 
Within this approximation, the time derivative of Eq.~\eqref{msdrm} reads
\begin{eqnarray}
\odif{R}{t} 
&=& \frac{4\sqrt{2}}{\pi} \int^{2-\epsilon}_{-\epsilon} d\lambda \sqrt{(\lambda + \epsilon)} \mathrm{e}^{-2 \lambda t} \nonumber \\
&=& \frac{2}{\pi} \mathrm{e}^{2 \epsilon t} t^{-3/2} 
\int^{4t}_{0} \sqrt{s} \mathrm{e}^{- s} ds. 
\end{eqnarray}
The integral is the incomplete gamma function $\gamma(3/2,4t)$ and rapidly converges to $\Gamma(3/2) =\sqrt{\pi}/2$ for $t \gg 1$. 
On the other hand, for $t \ll \tau_{\beta}$, where $\tau_{\beta} = 1/\epsilon$ is the $\beta$-relaxation time, the exponential factor $\mathrm{e}^{2\epsilon t}$ rapidly converges to 1. 
Therefore, we obtain $\odif{R}{t} = \frac{1}{\sqrt{\pi}} t^{-3/2}$ for $1 \ll t \ll \tau_{\beta}$. 
This implies the following power-law behavior 
\begin{eqnarray}
R(t) = R_{\infty} - \frac{2}{\sqrt{\pi t}}, \label{msdbeta}
\end{eqnarray}
for $1 \ll t \ll \tau_{\beta}$ and an exponential divergence for $t \gg \tau_{\beta}$.  
This also means that
\begin{eqnarray}
\frac{F_s(k,t)}{F_{s,\infty}(k)} = 1 + \frac{k^2}{\sqrt{\pi  t}}, \label{fsktbeta}
\end{eqnarray}
for $1 \ll t \ll \tau_{\beta}$ and $F_s(k,t)$ rapidly converges to 0 when $t \gg \tau_{\beta}$ due to the exponential divergence of $R(t)$.
In Fig.~\ref{fig:asymptotic}(a), we compare the asymptotic formula Eq.~\eqref{fsktbeta} with the numerical data presented in Fig.~\ref{fig:modelfskt}. 
Clearly, Eq.~\eqref{fsktbeta} works perfectly for the numerical data, meaning that the power-law decay in the $\beta$-relaxation regime $F_s(k,t) - F_{s,\infty}(k) \propto t^{-a}$ takes place with the exponent $a = 1/2$ in the schematic model. 
We also measure the $\beta$-relaxation times numerically as $F_s(k,\tau_\beta) = F_{s.\infty}(k)$, as shown in Fig.~\ref{fig:asymptotic}(b). 
Consistent with the asymptotic analysis, $\tau_{\beta}$ diverges as  $\epsilon^{-1}$. 
We also show the $\alpha$-relaxation time, defined by $F_s(k,\tau_\alpha) = e^{-1}$, 
and plot the ratio $\tau_{\alpha}/\tau_{\beta}$ in Fig.~\ref{fig:asymptotic}(b). 
This ratio converges to about 10 as $\epsilon \to 0$ meaning that the $\alpha$-relaxation within this model simply tracks the $\beta$-relaxation.

Now, we perform a similar asymptotic analysis for the dynamic susceptibility. 
The second derivative of $\chi_{R,{\rm iso}}(t)$ can be calculated in the same way as the first derivative of $R(t)$. 
Focusing on the $\beta$-relaxation regime and applying the same approximation for the exponential function and incomplete gamma function, we obtain 
\begin{eqnarray}
\oodif{\chi_{R,{\rm iso}}}{t} 
&=& \frac{32\sqrt{2}}{\pi} \int^{2-\epsilon}_{-\epsilon} d\lambda \sqrt{(\lambda + \epsilon)} \left( - \mathrm{e}^{-2 \lambda t} + 2 \mathrm{e}^{-4 \lambda t} \right) \nonumber \\
&=& \frac{-8 + 4 \sqrt{2}}{\sqrt{\pi}} t^{-3/2}, 
\end{eqnarray}
in the time range $1 \ll t \ll \tau_{\beta}$. 
This implies 
\begin{eqnarray}
&& \chi_{R,{\rm iso}}(t) = \frac{32 - 16 \sqrt{2}}{\sqrt{\pi}} \sqrt{t}, \\ 
&& \chi_{4,{\rm iso}}(k,t) = k^4 \mathrm{e}^{-k^2 R_{\infty}} \frac{8 - 4 \sqrt{2}}{\sqrt{\pi}} \sqrt{t}, \label{chi4beta}
\end{eqnarray}
in the early $\beta$-relaxation regime. 
These asymptotic expressions are included in Figs.~\ref{fig:modelmsd}(b) and \ref{fig:modelfskt}(b): they perfectly describe the scaling of the numerical data. 
Therefore, the four-point dynamic susceptibility follows $\chi_{4,{\rm iso}}(k,t) \propto t^{a'}$ in the early $\beta$-regime with $a' = 1/2$. 
This also means that the dynamic susceptibility at the $\beta$-relaxation time diverges as $\chi_{4,{\rm iso}}(k,\tau_{\beta}) \propto \epsilon^{-1/2}$.

In summary, within the schematic model, the power-law exponents for $F_s(k,t)$ and $\chi_{4,{\rm iso}}(k,t)$ in the early $\beta$-regime are the same $a = a' = 1/2$. 
We now compare this result with the predictions of MCT. 
Within the so-called inhomogeneous MCT, the dynamic susceptibility is computed as the response to a weak, spatially modulated perturbation~\cite{Biroli2006}. 
\red{Successive studies on the terms contributing to $\chi_4$~\cite{Berthier2007a,Franz2011a,Rizzo_Voigtmann_2020} showed that $\chi_{4, {\rm iso}}^2 \sim \chi_4$, which means $\chi_4 \sim t^{2a}$ and $\chi_{4, {\rm iso}} \sim t^{a}$ in the early $\beta$-regime.}  
Therefore, the framework of the MCT predicts $a = a'$ \red{within the isoconfigurational ensemble}. 
Similarly, this framework predicts $\chi_{4, {\rm iso}}(k,\tau_{\beta}) \propto \epsilon^{-1/2}$, which is observed in the schematic model too. 
Therefore, the schematic model reproduces all these MCT predictions for the relaxation dynamics and dynamic heterogeneity in the early $\beta$-relaxation regime within the isoconfigurational ensemble.
Note that $a$ is different from the exponent $b$, which controls the late $\beta$-relaxation, \textit{i.e.}, the departure from the plateau~\cite{gotze_complex_2009}.
Within MCT, $a$ and $b$ follow the equation $\Lambda = \frac{\Gamma(1-a)^2}{\Gamma(1-2a)} = \frac{\Gamma(1+b)^2}{\Gamma(1+2b)}$, where $\Lambda$ is a system-dependent constant~\cite{gotze_complex_2009}.
For the $p$-spin spherical model with $p=3$ one finds $\Lambda = 1/2$ and thus $a \approx 0.395$~\cite{Crisanti1993}; therefore, this model has different exponents than the schematic model even though its spectrum follows the semi-circle law~\cite{cavagna_single_2003}. 
For hard spheres in $d=3$, $\Lambda \approx 0.735$ and thus $a \approx 0.312$~\cite{gotze_complex_2009}. 
Here, it is interesting to note that, within MCT, $a=1/2$ corresponds to $\Lambda \to 0$. 
In this limit, the exponent for the late $\beta$-relaxation diverges, $b \to \infty$, which means that the power-law behavior in the late $\beta$-regime is absent and the time scales of the $\alpha$ and $\beta$-relaxations become identical $\tau_{\alpha} \propto \tau_{\beta}$, which we exactly observed in our schematic model, too. 
This observation suggests that the simple schematic model discussed in this section might correspond to MCT in the special case $\Lambda \to 0$. 
However, we also note that $\Lambda < 1/2$ is usually observed for continuous transitions, which lack a two-step relaxation~\cite{gotze_complex_2009}.
This point requires further investigation. 

%%%%%%%%%%%%%%%%%%%%%%%%%%%%%%%%%%%%%%%%%%%%%%%%%%%%%%%%%%%%%%%%%%%%
\section{Langevin dynamics simulations}\label{sec:supercooled}
%%%%%%%%%%%%%%%%%%%%%%%%%%%%%%%%%%%%%%%%%%%%%%%%%%%%%%%%%%%%%%%%%%%%
In this section, we directly compare the predictions of the SSM to the results of overdamped Langevin dynamics computer simulations for a model supercooled mixture. Data production and analysis have been carried out using a reproducible workflow, which is deposited in the Zenodo public repository~\cite{zenodo}.

We study the ternary mixture introduced by Gutiérrez~\textit{et al.} in Ref.~\onlinecite{gutierrez_static_2015}. The model is composed of $N=1000$ point particles interacting with an inverse power potential $u(r)=\epsilon (\sigma_{\alpha\beta}/r)^{12} + c_{\alpha\beta}(r)$, where $\alpha$, $\beta = A, B, C$ are species indices. The correction term $c_{\alpha\beta}(r)$ ensures that the second derivative is continuous at the cutoff distance $r_c = 1.25\sigma_{\alpha\beta}$. Energies and distances are given in units of $\epsilon$ and $\sigma_{AA}$. More details can be found in the original paper as well as in Ref.~\onlinecite{ninarello_models_2017}. The system can been equilibrated around and even below the MCT crossover temperature using the swap Monte Carlo algorithm \cite{gutierrez_static_2015,ninarello_models_2017}.

The saddles of the system have been located in Ref.~\onlinecite{coslovich_localization_2019} using the eigenvector-following (EF) method \cite{wales2003energy}. This algorithm searches for a stationary point of prescribed order $n_u$ in the neighborhood of the initial equilibrium configuration. In Ref.~\onlinecite{coslovich_localization_2019}, the target value of $n_u$ for a given optimization was fixed to the number of unstable modes found in a neighboring ``quasi-saddle", located using a mean square force minimization~\cite{grigera_geometric_2002}. Full details about the protocol can be found in Ref.~\onlinecite{coslovich_localization_2019}. In the following, we will focus on saddles obtained from equilibrium configurations sampled at $T=0.35$ and $T=0.29\approx T_\textrm{MCT}$. For each temperature, we considered $30$ saddle configurations, tightly converged to values of the mean square force $W$ of order $10^{-12}$. To complement our analysis, we also located local minima of the potential energy using a standard conjugate gradient algorithm.

Starting from these stationary points, we carried out overdamped Langevin dynamics simulations at a run temperature $T_r$ to compute the correlation functions of interest. Except where otherwise noted, $T_r$ will be identical to the temperature $T$ at which the stationary points were sampled. Note that, strictly speaking, the system is out of equilibrium during our simulation. However, we only found minor differences between this out-of-equilibrium protocol and the results of simulations at equilibrium.  We integrated the equations of motion using a simple Ermak algorithm with a time step $\delta t=0.0001$. We checked that the resulting dynamic properties were compatible within error bars with those obtained with different time steps, $\delta t=0.0005$ and 0.00005. For each starting configuration, we carried out \red{$20$} independent simulations over a time scale comparable to one structural relaxation time. Each simulation used a different seed for the random number generator. The resulting isoconfigurational dynamic properties~\cite{Widmer-Cooper_Harrowell_Fynewever_2004} were then averaged over the full set of initial saddle configurations, \red{\textit{e.g.}, $R(t) = \langle \langle \hat{R}(t)\rangle\rangle_c$. We emphasize that this setup precisely corresponds to the one used in the SSM calculations.}

\begin{figure}
    \centering
    \includegraphics[scale=.95]{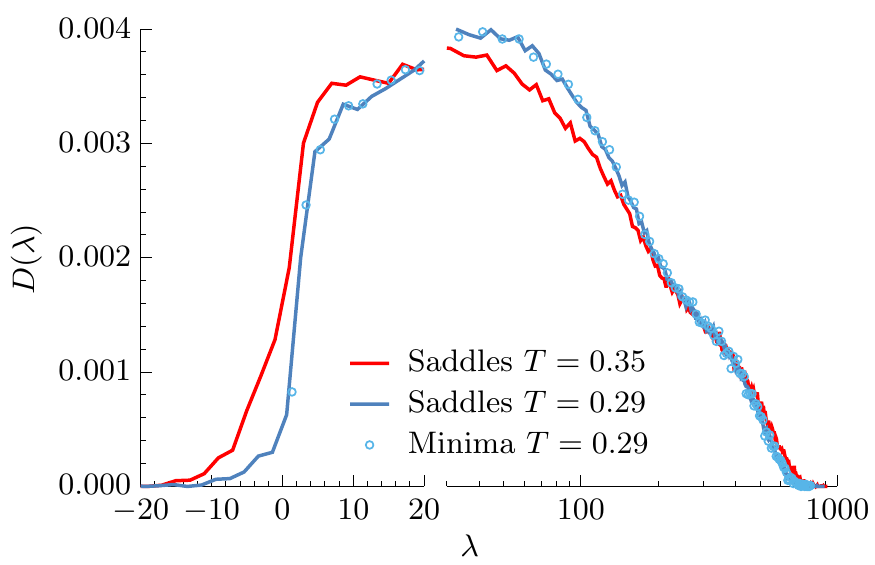}
    \caption{Spectrum of the Hessian $D(\lambda)$ for saddles sampled at $T=0.35$ and $T=0.29$, and for local minima sampled at $T=0.29$.} 
    \label{fig:vdos}
\end{figure}

To provide a reference for the following analysis, we show in Fig.~\ref{fig:vdos} the spectrum $D(\lambda)$ of the saddles sampled at the two temperatures of interest. Note that, above the MCT crossover, the unstable modes comprise both spatially localized and delocalized excitations~\cite{coslovich_localization_2019}. They can be distinguished, on average, by comparing their eigenvalue to the mobility edge $\lambda_e$: modes with $\lambda<\lambda_e$ and $\lambda_e < \lambda < 0$ are localized and delocalized, respectively, see Ref.~\onlinecite{coslovich_localization_2019} for further details. The mobility edge is $-4.6\pm 0.5$ at $T=0.35$ and nearly vanishes at $T=0.29$, at which almost all the unstable modes of a finite system are spatially localized. We also include the spectrum of the local minima sampled at $T=0.29$. At this temperature the stable branch of the saddle spectrum is practically indistinguishable from the one of the local minima.

\subsection{Mean square displacement}
\label{sec:sim_msd}

\begin{figure}[t]
 \begin{center}
  \includegraphics[scale=0.98]{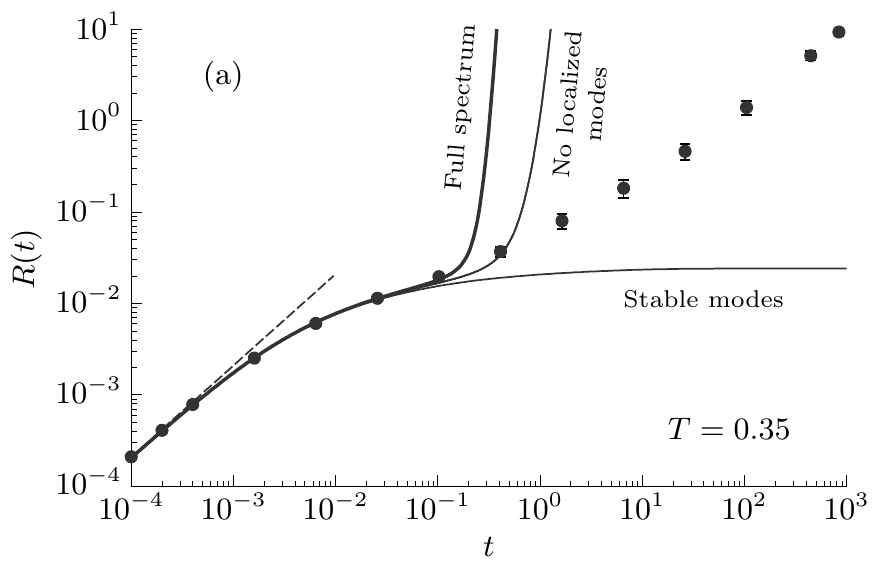}
  \includegraphics[scale=0.98]{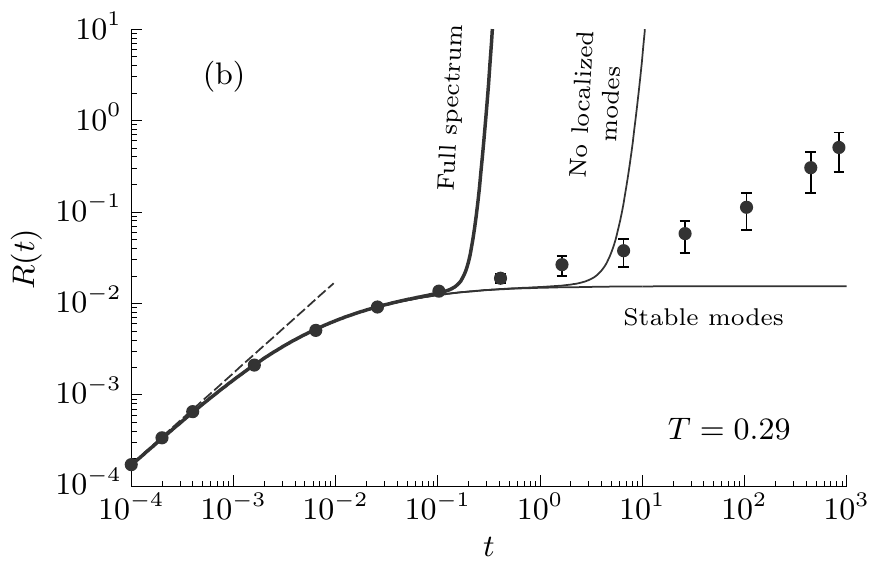}
  \caption{Mean square displacement $R(t)$ from the SSM (solid line) and from simulations (symbols) from saddles sampled at (a) $T=0.35$ and (b) $T= 0.29$. The dashes lines indicate the short-time, ballistic behavior.}
  \label{fig:msd}
 \end{center}
\end{figure}

We start by comparing the SSM predictions for the MSD with the numerical results of the Langevin dynamics simulations. Given the assumption of local harmonicity, the SSM predictions are only meaningful in the short-time and  $\beta$-relaxation time scale. In Fig.~\ref{fig:msd} we see that the agreement is perfect up to times of about 0.1, but it breaks down at longer times and the SSM solution diverges exponentially. As in previous simulation studies based on stochastic dynamics~\cite{Gleim1998}, we also do not observe a well-defined plateau in $R(t)$. Nonetheless, it is possible to define a $\beta$-relaxation time scale from the presence of an inflection in $R(t)$. By inspection of the figure, we see that the largest time $t^*\approx 0.1$ at which the SSM predictions and the simulation  agree, corresponds approximately with the inflection point. We thus conclude that the SSM provides an accurate description of the MSD in the early $\beta$-relaxation. \red{We also point out the behavior predicted by the SSM in this regime is not necessarily a critical one, i.e., power law. The shape of the correlation functions depends in general on the spectra and it is only in some special cases that the model predicts a critical approach to the plateau~\cite{cavagna_single_2003}}. 

If we remove the contribution of the unstable localized modes by restricting the integral in Eq.~\eqref{eqn:msd} to $\lambda > \lambda_e$, we find that at $T=0.35$ the SSM predictions track the numerical $R(t)$ over a slightly longer timescale, before eventually diverging at longer times, see Fig.~\ref{fig:msd}(a). Close inspection, however, shows that the agreement obtained through this empirical modification is qualitative at best, and that the theoretical curve is slightly below than the numerical one in this extended range of times. This discrepancy becomes more evident if we consider the saddles sampled at $T=0.29$, see Fig.~\ref{fig:msd}(b). At this temperature, this empirical correction leads to an average between two types of contributions: a fully frozen MSD profile, associated to saddles that do not possess delocalized unstable modes, and a few exponentially diverging contributions associated to residual localized unstable modes. Since the empirical correction does not lead to an improved agreement, we will not consider it further.

\subsection{Intermediate scattering functions}
\label{sec:sim_fkt}

\begin{figure}
    \centering
    \includegraphics[scale=0.98]{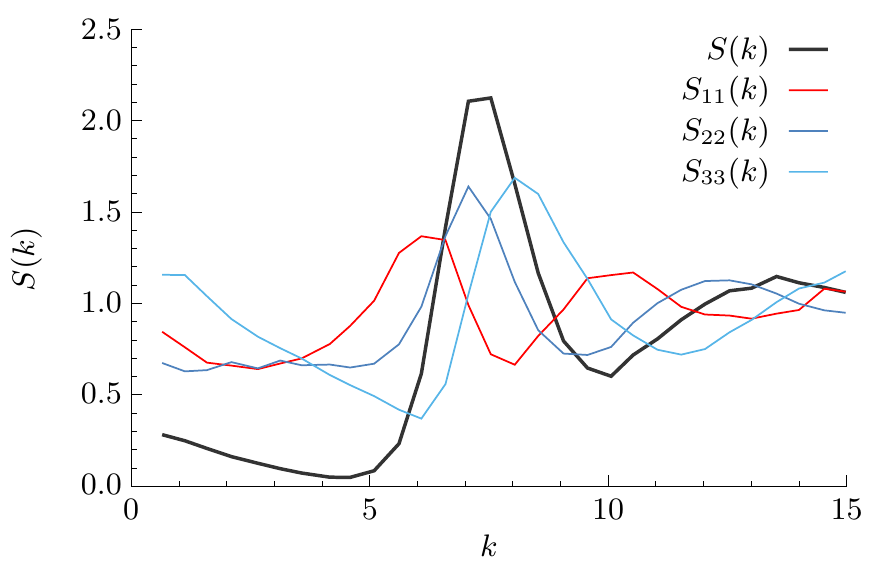}
    \caption{Static structure factor $S(k)$ (thick line) and partial structure factors $S_{\alpha\alpha}(k)$ (thin lines) from equilibrium configurations at $T=0.35$.}
    \label{fig:sk}
\end{figure}

We now investigate to what extent the SSM is able to capture correlations in space and time by analyzing the collective intermediate scattering function $F(k,t)$ and its self part $F_s(k,t)$. We will carry out the calculations at several wave-numbers $k=|\vec{k}|$. For each wave-number $k$, we calculated the self (collective) intermediate scattering functions by spherically averaging over 10 (100) wave-vectors with norm in the interval $[k-0.1, k+0.1]$. We used exactly the same set of wavevectors to compute the correlation functions from simulations and within the SSM. For reference, we show in Fig.~\ref{fig:sk} the total structure factor $S(k)$ and the partial structure factors $S_{\alpha\alpha}(k)$ obtained from the simulations at $T=0.35$. The first peak of $S(k)$ occurs around $k\approx 7.5$. We observe a slight increase of the $S(k)$ at small $k$, which is due to the contribution of the largest particles in the system (see $S_{33}(k)$). We found that the corresponding structure factors calculated from saddle configurations are practically indistinguishable from the equilibrium ones at a given $T$, in agreement with Ref.~\onlinecite{Shimada_Coslovich_Mizuno_Ikeda_2021}.

\begin{figure}[t]
 \begin{center}
  \includegraphics[scale=0.98]{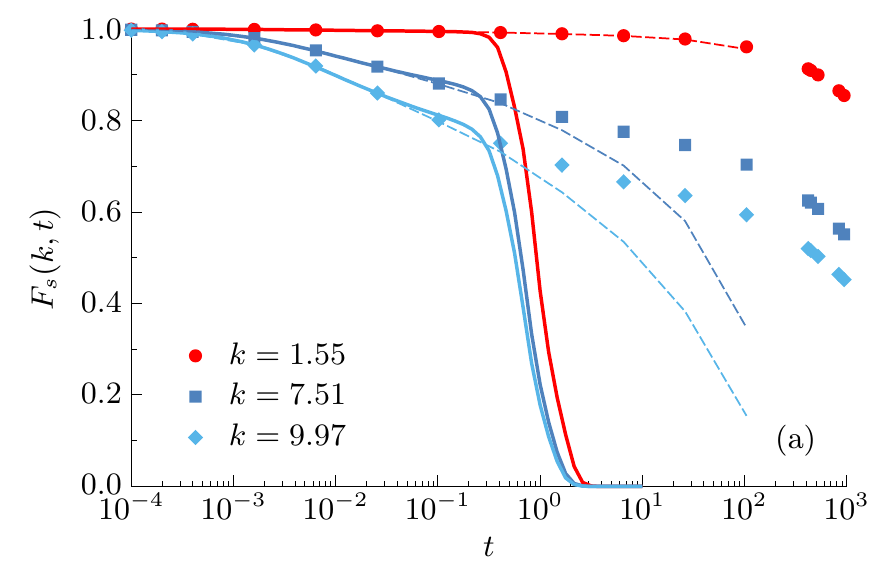}
  \includegraphics[scale=0.98]{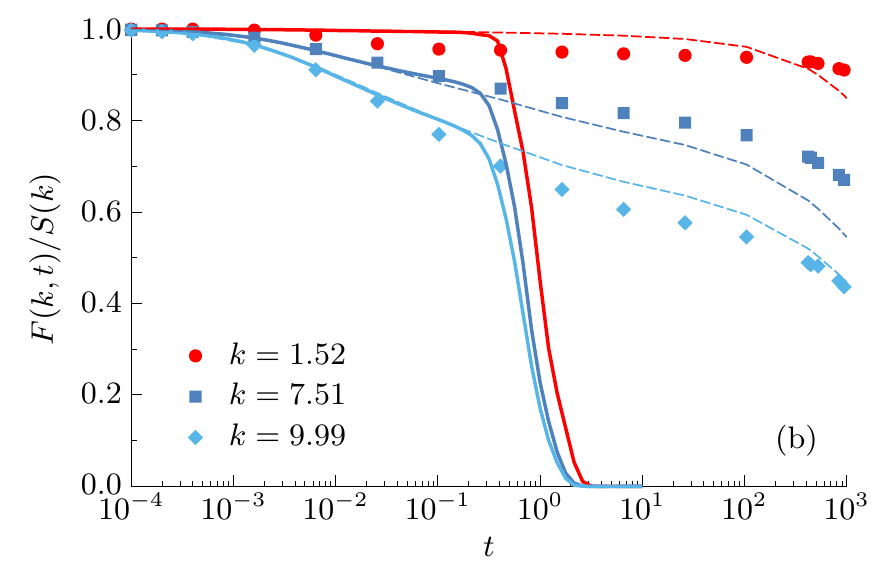}
  \caption{SSM predictions (solid lines) and simulation results (symbols) for (a) $F_s(k,t)$ and (b) $F(k,t)$ from saddles sampled at $T=0.29$. The chosen wave-number $k$ is indicated in the figure. The dashed lines in are the results of (a) Gaussian approximation, Eq.~\eqref{eq:gaussian_approx} and of (b) the Vineyard approximation Eq.~\eqref{eqn:vineyard}.}
  \label{fig:fkt}
 \end{center}
\end{figure}

We analyze $F_s(k,t)$ first (see Fig.~\ref{fig:fkt}(a)), focusing on three representative wave-numbers: $k=1.55$, $7.51$, and $9.97$. They correspond to the low-$k$ region, the first peak, and the first minimum of the total structure factor, respectively. We find that the SSM predictions agree almost perfectly with the simulation data in the early $\beta$-regime, as already found for the MSD, irrespective of the wave-number. The decay to zero of the correlation function at longer times is, of course, too rapid. We note that on the late $\beta$-regime, where the SSM breaks down, a simple Gaussian approximation
\begin{eqnarray}\label{eq:gaussian_approx}
F_s(k,t) = \mathrm{e}^{- k^2 R(t)/6}
\end{eqnarray}
works pretty well especially at large $k$.
We point out, however, that the Gaussian approximation is not ``predictive", because it requires some dynamic information, \textit{i.e.}, the mean square displacement, in the first place.
Moving on to the total correlation $F(k,t)$ (see Fig.~\ref{fig:fkt}(b)) and restricting again our analysis to short times, the agreement looks fair for $k$ close to the first peak of $S(k)$ but deteriorates at the other wave-vectors. The behavior at short times, for the three wave-numbers considered here, closely tracks the results of the Vineyard approximation Eq.~\eqref{eqn:vineyard}. Qualitatively, these results suggest that the SSM captures the single particle motion better than collective density fluctuations.

\begin{figure}[t]
 \begin{center}
  \includegraphics[scale=0.98]{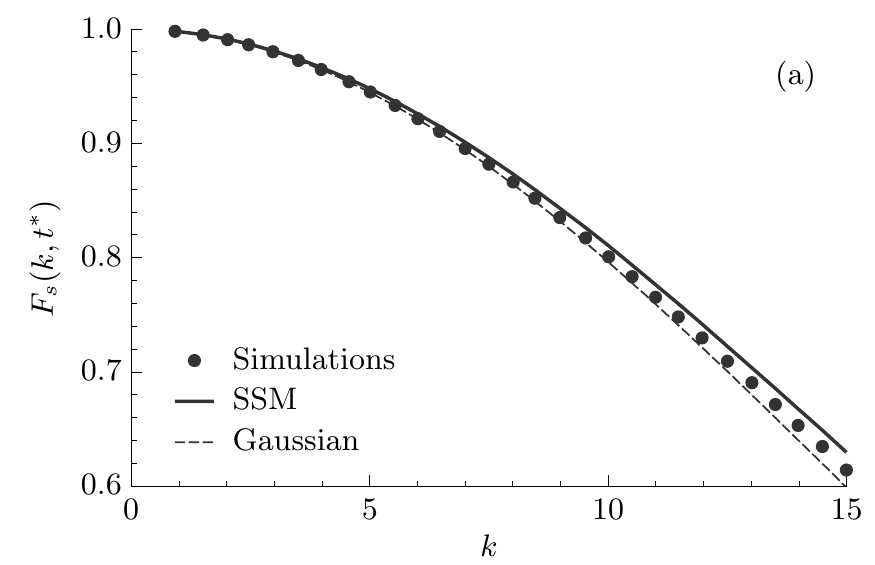}
  \includegraphics[scale=0.98]{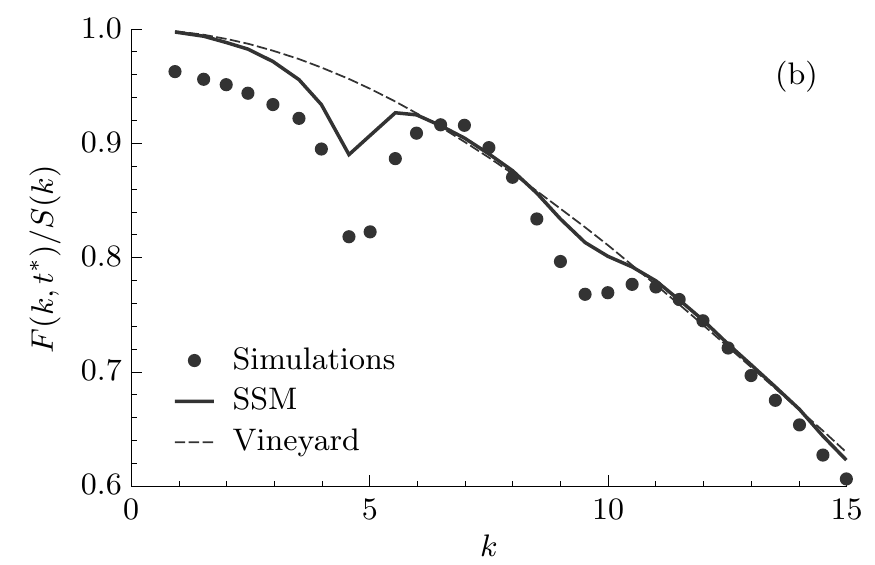}
  \caption{SSM predictions (solid lines) and simulation results (symbols) for (a) $f_s(k) = F_s(k,t^*)$ and (b) $f(k) = F(k,t^*)$ from saddles sampled at $T=0.29$. The dashed lines in (a) and (b) indicate the Gaussian approximation Eq.~\eqref{eq:gaussian_approx} and the Vineyard approximation Eq.~\eqref{eqn:vineyard}, respectively.}
  \label{fig:fkt_kdep}
 \end{center}
\end{figure}

To analyze this point more in-depth, we consider the $k$-dependence of the correlation functions at $t=t^*=0.1024$, which is approximately the largest time at which the MSDs from theory and simulations match well. Since $t^*$ is close to the inflection in $R(t)$, the functions $f(k)=F(k,t^*)/S(k)$ and $f_s(k) = F_s(k,t^*)$ are proxies to the corresponding non-ergodicity parameters, which measure the plateau height of the scattering functions in a dynamically arrested system.
Figure~\ref{fig:fkt_kdep} shows that the SSM captures well the Gaussian $k$-dependence of $f_s(k) \approx \exp{[- R(t^*) k^2 / 6]}$. The agreement at the level of $f(k)$ is \red{less satisfactory. 
The SSM qualitatively reproduces} the well-known peak of $f(k)$ in correspondence to the first peak of $S(k)$, while this feature is obviously missing in the Vineyard approximation $f(k)=f_s(k)$.
This suggests the existence of subtle correlations between the structure of the initial configuration and the eigenvectors, see also Sec.~\ref{sec:discussion}. 
However, we also see that the SSM overestimates $f(k)$ at small $k$ and that the maximum is slightly shifted. The agreement observed in Fig.~\ref{fig:fkt} for wave-numbers close to the first peak of $S(k)$ may therefore be partly coincidental. Our results show that quantitatively predicting the non-ergodicity parameters is a nontrivial task and, in retrospect, praise the ability of MCT \red{to account for these properties}~\cite{Kob_Nauroth_Sciortino_2002}.

\subsection{Dynamic susceptibility}
\label{sec:sim_chi4}

\begin{figure}[t]
 \begin{center}
  \includegraphics[scale=0.98]{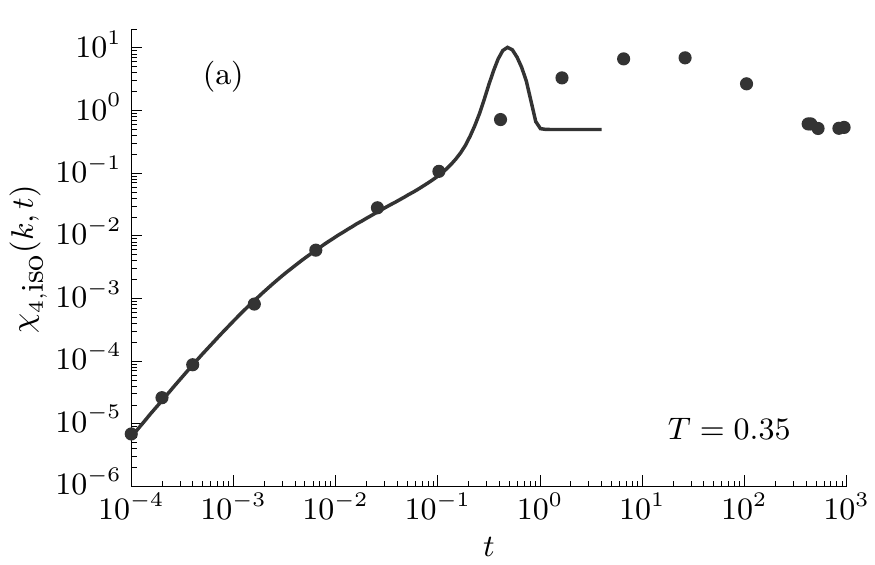}
  \includegraphics[scale=0.98]{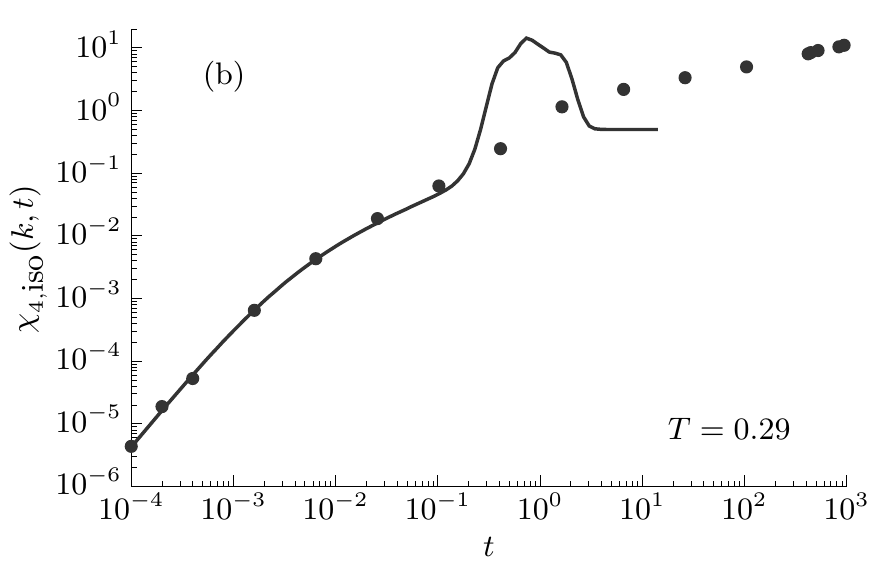}
  \caption{SSM predictions (solid line) and simulation results \red{(symbols)} for the isoconfigurational dynamic susceptibility $\chi_{4,\textrm{iso}}(k,t)$ from saddles sampled at (a) $T=0.35$ and (b) $T=0.29$.}
  \label{fig:chi4}
 \end{center}
\end{figure}

We now check the SSM predictions for the dynamic fluctuations of the single-particle dynamics. In Fig.~\ref{fig:chi4}, we show the dynamic susceptibility $\chi_{4,{\rm iso}}(k,t)$ calculated for a single wave-vector of norm $k=7.164$.
To match the SSM calculation, we computed $\chi_{4,{\rm iso}}$ within the isoconfigurational ensemble~\cite{Widmer-Cooper_Harrowell_Fynewever_2004}
\begin{eqnarray}
\chi_{4,{\rm iso}}(k,t) = N [\langle\langle \hat{F}_s^2(k,t)\rangle - \langle \hat{F}_s(k,t)\rangle^2 \rangle_c],
\end{eqnarray}
\red{where $\hat{F}_s(k,t)$ is the self intermediate scattering function calculated starting from a single configuration and for a single realization of the noise. As already mentioned in Sec.~\ref{sec:ssm}, the full dynamic susceptibility $\chi_4(k,t)$ contains an additional term associated to sample-to-sample fluctuations. We found that this term is negligible in the time range over which the SSM predictions work well (not shown). Therefore, we will not consider it further}.

The results of these calculations are shown in Fig.~\ref{fig:chi4}. Of course, the peak of the dynamic susceptibility predicted by SSM occurs at times shorter than the maximum observed in the simulations. The peak is also too high and sharp, which reflects the rapid decorrelation due to the unstable modes~\footnote{Note that at $T=0.29$,
the simulation data have not reached yet the maximum, since in this work we focus only on the $\beta$-regime.}. However, the agreement is again very good in the early $\beta$-regime. Thus, the SSM captures both the average single-particle dynamics and its fluctuations very well in this time range.

\subsection{Results for local minima}
\label{sec:org2d494c9}
Our analysis so far has shown that the SSM works very well at short times but that the agreement rapidly deteriorates on longer timescales, when the harmonic approximation inherent in the SSM breaks down. In an attempt to study a regime where the harmonic approximation should be obeyed over a longer time interval, we analyze the dynamics close to local minima of the potential energy surface. We consider local minima sampled at $T=0.29$ and simulate the system at a run temperature $T_r=0.02 \ll T$ using a timestep $\delta=0.0005$.

\begin{figure}
    \centering
    \includegraphics[scale=.95]{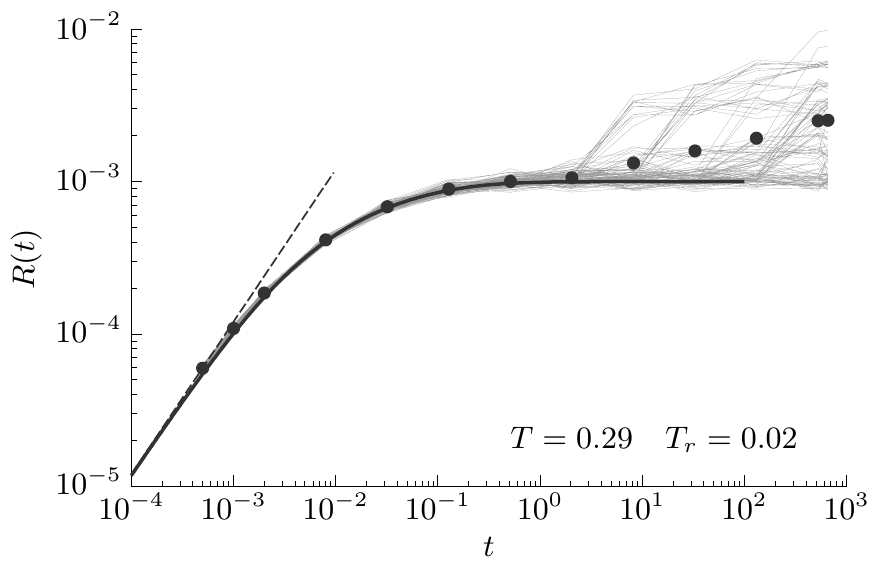}
    \caption{SSM prediction (solid line) and simulation results \red{(symbols)} for the mean square displacement $R(t)$ from local minima sampled at $T=0.29$ with a run temperature $T_r=0.02$. The dashed line indicates the short-time, ballistic behavior.}
    \label{fig:msd_min}
\end{figure}

In Fig.~\ref{fig:msd_min}, we show the MSD obtained with this setup. Because of the absence of unstable modes, the dynamics of the SSM is now completely frozen at long times. Compared to our previous analysis, the agreement with the SSM now stretches by one additional order of magnitude and is very good up to about $t\approx 1$. However, we see that in this time regime, the dynamics is highly heterogeneous and some samples display small scale rearrangements, associated to transitions between close-by minima. Thus, the SSM holds well over long times for samples that have not relaxed, but it is obviously unable to capture these rare dynamic transitions.

We also analyzed the $k$-dependence of $F_s(k,t^*)$ and $F(k,t^*)$, obtained starting from local minima (not shown). We found that the spatial structure of single-particle relaxation on short time scale was perfectly reproduced, but appreciable deviations persisted for the collective density fluctuations \red{at wave-vectors around and below the position of the first peak of $S(k)$}, which suggests that the subtle anharmonicities at short times play an important role for the collective density fluctuations. This point needs further investigation. 

\subsection{Discussion}
\label{sec:discussion}

In an effort to find ways to improve the model, we now analyze in more detail the connection between the relaxation dynamics and the eigenmodes. In particular, we show that the spatial structure of the unstable modes carry relevant information about the dynamics even beyond the $\beta$-regime.

We consider the isoconfigurational square mobility of particles $\mu^2_i(t) = \langle | \vec{r}_i(t) - \vec{r}_i(0)|^2\rangle$ and we compute its correlation with the average norm square of selected eigenvectors, $E_i^2 = 1/n \sum_\alpha |\vec{e}_{\alpha,i}|^2$, where $n$ is the number of selected modes. We consider separately the subset of unstable eigenvectors, $\lambda_\alpha<0$ and the subset of soft stable modes $0<\lambda_\alpha<2.0$. We then compute the standard Pearson correlation coefficient $K_P(t)$ and the Spearman correlation coefficient $K_S(t)$, defined as the Pearson correlation coefficients between the ranks of the sorted variables. This procedure is common in the analysis of the correlation between structural order metrics and local dynamics
~\cite{hocky_correlation_2014,Paret_Jack_Coslovich_2020,Boattini_Marin-Aguilar_Mitra_Foffi_Smallenburg_Filion_2020,Boattini_Smallenburg_Filion_2021}.

\begin{figure}[t]
 \begin{center}
  \includegraphics[scale=0.98]{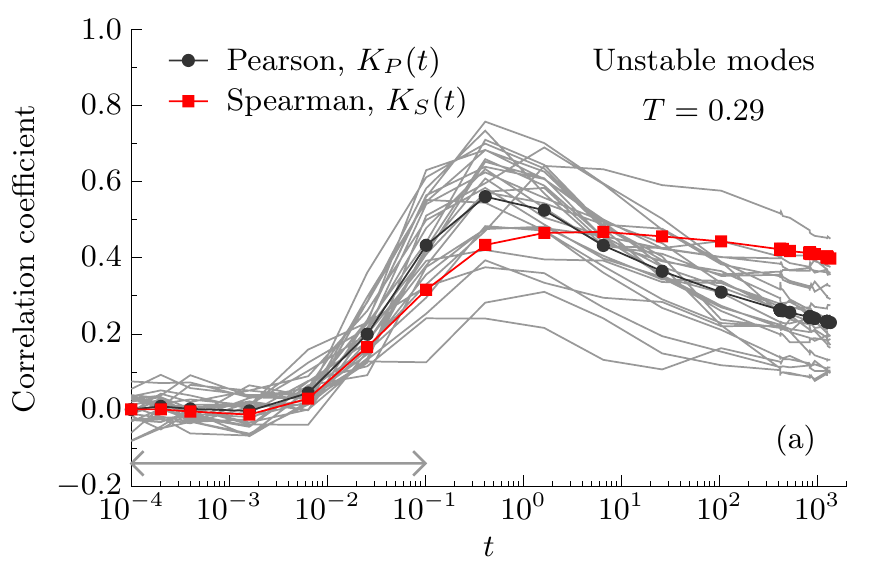}
  \includegraphics[scale=0.98]{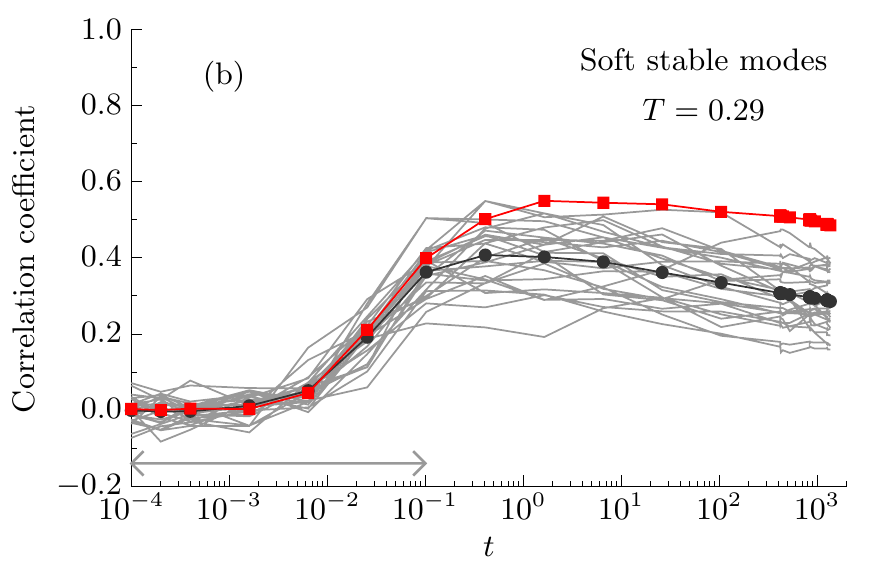}
  \caption{Correlation coefficients $K_P(t)$ and $K_S(t)$ between the isoconfigurational square mobility $\mu_i^2(t)$ and the average square displacements $E_i^2$ on (a) the unstable modes and (b) soft modes ($0<\lambda_\alpha<2.0$) of saddles sampled at $T=0.29$. The thin gray lines correspond to correlation coefficients $K_P(t)$ calculated for individual configurations, while the symbols indicate averages of the correlation coefficients over all the configurations. The horizontal line indicates the time range over which the SSM works well.}
  \label{fig:correlation}
 \end{center}
\end{figure}

In Fig.~\ref{fig:correlation} we show $K_P$ and $K_S$ as a function of time. The analysis is carried out for saddles sampled at $T=0.29$ both at the level of individual configurations (thin lines) and averaging over all the configurations (symbols). We see that beyond the time scale $t^*$, up to which the SSM works well, the correlation coefficients steadily increase and reach a broad maximum at about 0.5 before slowly decreasing on approaching the structural relaxation time. Values of $K_S$ of about 0.5 are indicative of a significant correlation between unstable modes and local dynamics~\cite{hocky_correlation_2014}. Similar correlations are found with the soft stable modes, see Fig.~\ref{fig:correlation}(b), in agreement with Ref.~\onlinecite{jack_information-theoretic_2014}. Note that we did not average $E_i^2$ over the neighboring particles, as was done in previous work~\cite{hocky_correlation_2014,Paret_Jack_Coslovich_2020,Boattini_Marin-Aguilar_Mitra_Foffi_Smallenburg_Filion_2020,Boattini_Smallenburg_Filion_2021} to further increase the correlation at long times. We conclude that the unstable modes are predictive of the local dynamics also in the late $\beta$-regime, but the SSM is currently not able to exploit this information. 

One obvious unphysical aspect of the model is that the system rolls away without bounds along the unstable modes of the saddle, while in the actual dynamics it will stop and fluctuate at the bottom of some neighboring local minimum. To partly correct this issue, anharmonicity should be taken into account. This could be done \textit{ad hoc} by suppressing contributions from unstable modes when the value of $K(\lambda_\alpha,t)$ (see Eq.~\eqref{eqn:S_and_K}) exceeds a threshold. Preliminary attempts along these lines, however, did not lead to an improved agreement with the simulations. Alternatively, once the value of $K(\lambda_\alpha,t)$ of a given unstable mode reaches a threshold, one could replace the exponential divergence with a diffusive contribution proportional to the participation ratio and an appropriate diffusion constant. It would be interesting to develop a more systematic approach to account for anharmonicity, similar to what was done long ago for instantaneous normal modes~\cite{Kramer_Buchner_Dorfmuller_1998}, and to establish connections with alternative approaches to the $\beta$-relaxation dynamics, such as the stochastic $\beta$-relaxation model~\cite{Rizzo_2016,Rizzo_Voigtmann_2020}.

\section{Conclusions}

In this work, we studied the dynamics of supercooled liquids starting from saddle configurations using both numerical simulations and a simple theoretical model, first introduced by Cavagna \textit{et al.}~\cite{cavagna_single_2003}.

First, we extended the model to calculate various dynamical quantities within the harmonic approximation. In particular, we obtained predictions for the self and collective intermediate scattering functions as well as for the four-point dynamical susceptibility \red{in the isoconfigurational ensemble}. The obtained formulas allows one to calculate these dynamical quantities using the eigenvalues and eigenmodes at the saddle only. We note that it is easy to extend the model to compute these quantities from equilibrium configurations in the neighborhood of the saddle.

We then introduced a schematic model that assumes that the eigenmodes are randomly distributed and that the eigenvalues follow the semi-circle law, as in several mean-field spin glass models. 
In the schematic model, the dynamical quantities can be written as simple integrals and their asymptotic behaviors can be \red{calculated} analytically.
On approaching the dynamical transition, at which the unstable support of the spectrum vanishes, all the dynamic observables display power-law behavior in the $\beta$-regime with identical exponents, which is consistent with the predictions of MCT and its inhomogeneous extension within the isoconfigurational ensemble.
The power-law scaling of the dynamic susceptibility is identical to the one predicted by the much more complex setting of inhomogeneous MCT.
However, on a longer time scale, the schematic model exhibits a very rapid relaxation and therefore the $\beta$- and $\alpha$-relaxation times scale identically, in sharp contrast to the predictions of MCT and to the actual dynamics in the supercooled liquids. 

We performed overdamped Langevin simulations for a supercooled ternary mixture equilibrated close to the MCT crossover temperature and assessed the theoretical predictions of the SSM using actual saddles as input.
The agreement in the early $\beta$-regime is very good for the single particle dynamic properties, including the 4-point dynamic susceptibility, but only qualitative for the relaxation of collective density fluctuations.
We conclude that the model predictions are fair, but their current range of validity is too limited to be relevant for the structural relaxation of supercooled liquids.

Nonetheless, we think there is room for improvement.
In particular, on the time scale on which the SSM predictions break down, the unstable eigenmodes are still significantly correlated with the local dynamics and they remain so up to times of the order of the structural relaxation time. 
This indicates that the SSM may be largely improved by taking into account anharmonic effects or through corrections that better account for the spatial structure of the unstable modes. 
This might lead to a predictive, first-principles theoretical model of the supercooled liquid dynamics up to time scales 
comparable to the structural relaxation time.

\section*{Author's contributions}
Both authors contributed equally to this work.

\section*{Acknowledgments}
DC acknowledges support as a JSPS International Research Fellow.
AI acknowledges support by JSPS KAKENHI Grants No.~18H05225, 19H01812, 20H01868, 20H00128. 

\section*{Data availability}
The data and worflow necessary to reproduce the findings of this study are openly available in the Zenodo data repository at \url{https://doi.org/10.5281/zenodo.5791675}.

%%%%%%%%%%%%%%%%%%%%%%%%%%%%%%%%%%%%%%%%%%%%%%%%%%%%%%%%%%%%%%%%%%%%
\appendix
%%%%%%%%%%%%%%%%%%%%%%%%%%%%%%%%%%%%%%%%%%%%%%%%%%%%%%%%%%%%%%%%%%%%

%%%%%%%%%%%%%%%%%%%%%%%%%%%%%%%%%%%%%%%%%%%%%%%%%%%%%%%%%%%%%%%%%%%%
\section{Derivation of Eqs.~\eqref{fktresult} and \eqref{chi4result}}\label{sec:appendix1}
%%%%%%%%%%%%%%%%%%%%%%%%%%%%%%%%%%%%%%%%%%%%%%%%%%%%%%%%%%%%%%%%%%%%

To calculate the wave-vector dependent quantities, it is useful to introduce the Fourier transform of the solution Eq.~\eqref{fpsol}: 
\begin{eqnarray}
\int d \bm{x} \mathrm{e}^{i \bm{\xi} \cdot \bm{x} } P(\bm{x},t) = \mathrm{e}^{- \frac{1}{2} \bm{\xi} \cdot \mat{S}(t) \cdot \bm {\xi}}, \label{ft}
\end{eqnarray}
where $\bm{\xi}$ is a wave vector in the $dN$-dimensional configuration space.

To calculate $\ave{\hat{F}_s(k,t)}$, we need to calculate $\ave{\mathrm{e}^{i \vec{k} \cdot \vec{x}_i(t)}}$ and $\ave{\mathrm{e}^{-i \vec{k} \cdot \vec{x}_i(t)}}$. 
This can be done by introducing the $dN$-dimensional wave-vector $\bm{\xi}_i$ in which only the particle $i$ has a non-zero component equal to $\vec{k}$: $\bm{\xi}_i \equiv (..., \vec{0}, \vec{k}, \vec{0}, ...)$.
Then using Eq.~\eqref{ft} with $\bm{\xi}_i$, we obtain 
\begin{eqnarray}
\ave{\mathrm{e}^{i \vec{k} \cdot \vec{x}_i(t)}} = \mathrm{e}^{- \frac{1}{2} \bm{\xi}_i \cdot \mat{S}(t) \cdot \bm{\xi}_i}
= \mathrm{e}^{- \frac{T}{2} \sum_{\alpha} K(\lambda_{\alpha},t) (\vec{k} \cdot \vec{e}_{\alpha,i})^2}, 
\end{eqnarray}
which lead to the expression for $\ave{\hat{F}_s(k,t)}$ in Eq.~\eqref{fktresult}. 
For $\ave{\hat{F}(k,t)}$, we first transform the definition into
\begin{eqnarray}
\ave{\hat{F}(\vec{k},t)} = \frac{1}{N} \sum_{ij} \mathrm{e}^{i \vec{k} \cdot (\vec{r}_i - \vec{r}_j)} \Ave{ \mathrm{e}^{i \vec{k} \cdot \vec{x}_i(t)}},  
\end{eqnarray}
and then perform the same calculation as the self part, which gives the expression for $\ave{\hat{F}(k,t)}$ in Eq.~\eqref{fktresult}. 

For $ \ave{\hat{\chi}_{4,{\rm iso}}(\vec{k},t)}$, we have to calculate the average $\Ave{\left( \frac{1}{N} \sum_i \cos (\vec{k} \cdot \vec{x}_i(t)) \right)^2}$.
This consists of the contributions from particle pairs, $\Ave{ \mathrm{e}^{i \vec{k} \cdot (\vec{x}_i(t) + \vec{x}_j(t))}}$ and $\Ave{ \mathrm{e}^{i \vec{k} \cdot (\vec{x}_i(t) - \vec{x}_j(t))}}$. 
When $i\neq j$, $\Ave{ \mathrm{e}^{i \vec{k} \cdot (\vec{x}_i(t) + \vec{x}_j(t))}}$ can be calculated by introducing the wave vectors $\bm{\xi}_{ij}$ in which only the particles $i$ and $j$ parts have non-zero components: $\bm{\xi}_{ij} \equiv (...,\vec{0}, \vec{k},\vec{0}, ..., \vec{0}, \vec{k}, \vec{0}, ...)$. 
Then, we obtain 
\begin{eqnarray}
\ave{ \mathrm{e}^{i \vec{k} \cdot (\vec{x}_i(t) + \vec{x}_j(t))}} &=& \mathrm{e}^{- \frac{1}{2} \bm{\xi}_{ij} \cdot \mat{S}(t) \cdot \bm{\xi}_{ij}} \nonumber \\
&=& \mathrm{e}^{- \frac{T}{2} \sum_{\alpha} K(\lambda_{\alpha},t) (\vec{k} \cdot (\vec{e}_{\alpha,i} + \vec{e}_{\alpha,j}))^2}. 
\end{eqnarray}
Similarly, we can calculate $\Ave{ \mathrm{e}^{i \vec{k} \cdot (\vec{x}_i(t) - \vec{x}_j(t))}}$ by introducing $\bm{\xi}_{ij} \equiv (...,\vec{0}, \vec{k},\vec{0}, ..., \vec{0},-\vec{k}, \vec{0}, ...)$, and we obtain
\begin{eqnarray}
\ave{ \mathrm{e}^{i \vec{k} \cdot (\vec{x}_i(t) - \vec{x}_j(t))}} &=& \mathrm{e}^{- \frac{1}{2} \bm{\xi}_{ij} \cdot \mat{S}(t) \cdot \bm{\xi}_{ij}} \nonumber \\
&=& \mathrm{e}^{- \frac{T}{2} \sum_{\alpha} K(\lambda_{\alpha},t) (\vec{k} \cdot (\vec{e}_{\alpha,i} - \vec{e}_{\alpha,j}))^2}.
\end{eqnarray}
We can do similar calculations for the case $i = j$. 
Then, summing all the terms and using Eq.~\eqref{fktresult}, we obtain the expression for $ \ave{\hat{\chi}_{4,{\rm iso}}(\vec{k},t)}$ in Eq.~\eqref{chi4result}

%%%%%%%%%%%%%%%%%%%%%%%%%%%%%%%%%%%%%%%%%%%%%%%%%%%%%%%%%%%%%%%%%%%%
\section{Derivation of Eq.~\eqref{chi4rdm}}\label{sec:appendix2}
%%%%%%%%%%%%%%%%%%%%%%%%%%%%%%%%%%%%%%%%%%%%%%%%%%%%%%%%%%%%%%%%%%%%

Here, we calculate $\chi_{4,{\rm iso}}(k,t)$ in the schematic model. 
To this end, we split $\chi_{4,{\rm iso}}(k,t)$ into the self and distinct parts as $\chi_{4,{\rm iso}}(k,t) = \chi_{4,{\rm iso},{\rm self}}(k,t) + \chi_{4,{\rm iso},{\rm dist}}(k,t)$, where the self part is the contribution from $i=j$ terms in Eq.~\eqref{chi4result} and the distinct part is from $i \neq j$ terms.
The self part can be calculated in the same way as $F_s(k,t)$: 
\begin{widetext}
\begin{eqnarray}
\chi_{4,{\rm iso},\mathrm{self}}(k,t) &=& \frac{1}{N}\sum_i \prod_{\alpha} \int d\lambda_{\alpha} \rho(\lambda_{\alpha}) \int de_{\alpha,i} f(e_{\alpha,i}) 
\left[ \frac{1}{2} + \frac{1}{2} \mathrm{e}^{- 2 K(\lambda_{\alpha},t) k^2 e_{\alpha,i}^2}
-\mathrm{e}^{- K(\lambda_{\alpha},t) k^2 e_{\alpha,i}^2} \right] \nonumber \\
&=& \frac{1}{2} + \frac{1}{2} \left[ \int d\lambda \rho(\lambda) \left(1 + \frac{4k^2}{N} K(\lambda,t) \right)^{-1/2} \right]^N - \left[ \int d\lambda \rho(\lambda) \left(1 + \frac{2k^2}{N} K(\lambda,t) \right)^{-1/2} \right]^N 
= \frac{1}{2} (1 - \mathrm{e}^{- k^2 R(t)})^2, \nonumber \\
\end{eqnarray}
where we took the limit $N \to \infty$ in the final line. 
The distinct part consists of three contributions characterized by $(k (e_{\alpha,i} + e_{\alpha,j}))^2$, $(k (e_{\alpha,i} - e_{\alpha,j}))^2$, and $(k e_{\alpha,i})^2 + (k e_{\alpha,j})^2$, respectively. 
For each term, we obtain the following results: 
\begin{eqnarray}
&& \prod_{\alpha} \int d\lambda_{\alpha} \rho(\lambda_{\alpha}) \int de_{\alpha,i} f(e_{\alpha,i}) \int de_{\alpha,j} f(e_{\alpha,j}) \mathrm{e}^{- \frac{1}{2}  K(\lambda_{\alpha},t) (k (e_{\alpha,i} + e_{\alpha,j}))^2} = \left[ \int d\lambda \rho(\lambda) \left( 1 + \frac{2k^2}{N}K(\lambda,t) \right)^{-1/2} \right]^N, \nonumber \\
&& \prod_{\alpha} \int d\lambda_{\alpha} \rho(\lambda_{\alpha}) \int de_{\alpha,i} f(e_{\alpha,i}) \int de_{\alpha,j} f(e_{\alpha,j}) \mathrm{e}^{- \frac{1}{2}  K(\lambda_{\alpha},t) (e_{\alpha,i} - e_{\alpha,j}))^2} = \left[ \int d\lambda \rho(\lambda) \left( 1 + \frac{2k^2}{N}K(\lambda,t) \right)^{-1/2} \right]^N, \nonumber \\
&& \prod_{\alpha} \int d\lambda_{\alpha} \rho(\lambda_{\alpha}) \int de_{\alpha,i} f(e_{\alpha,i}) \int de_{\alpha,j} f(e_{\alpha,j}) \mathrm{e}^{- \frac{1}{2}  K(\lambda_{\alpha},t) (k e_{\alpha,i})^2 + (k e_{\alpha,j}))^2} = \left[ \int d\lambda \rho(\lambda) \left( 1 + \frac{k^2}{N}K(\lambda,t) \right)^{-1} \right]^N. \nonumber 
\end{eqnarray}
Gathering all terms, we obtain
\begin{eqnarray}
\chi_{4,{\rm iso},\mathrm{dist}}(k,t) &=& (N-1) \left\{ 
\left[ \int d\lambda \rho(\lambda) \left( 1 + \frac{2k^2}{N}K(\lambda,t) \right)^{-1/2} \right]^N 
- \left[ \int d\lambda \rho(\lambda) \left( 1 + \frac{k^2}{N} K(\lambda,t) \right)^{-1} \right]^N \right\} \nonumber \\
&=& (N-1) \left\{ \left[1 - \frac{k^2 R(t)}{N} + \frac{3k^4 \chi_R(t)}{4N^2} + O(N^{-3})\right]^N
- \left[1 - \frac{k^2 R(t)}{N} + \frac{k^4 \chi_{R,{\rm iso}}(t)}{2N^2} + O(N^{-3})\right]^N
\right\} \nonumber \\
&=& \frac{1}{4} k^4 \chi_{R,{\rm iso}}(t) \mathrm{e}^{- k^2 R(t)}, 
\end{eqnarray}
\end{widetext}
where again we took $N \to \infty$ in the final line. 
Note that the leading order contributions in the curly brackets precisely vanish, and only the second leading order terms remain. 
Summing the self and distinct parts, we obtain the expression of $\chi_{4,{\rm iso}}(k,t)$ in Eq.~\eqref{chi4rdm}.

\bibliography{single_saddle}

\begin{thebibliography}{58}%
\makeatletter
\providecommand \@ifxundefined [1]{%
 \@ifx{#1\undefined}
}%
\providecommand \@ifnum [1]{%
 \ifnum #1\expandafter \@firstoftwo
 \else \expandafter \@secondoftwo
 \fi
}%
\providecommand \@ifx [1]{%
 \ifx #1\expandafter \@firstoftwo
 \else \expandafter \@secondoftwo
 \fi
}%
\providecommand \natexlab [1]{#1}%
\providecommand \enquote  [1]{``#1''}%
\providecommand \bibnamefont  [1]{#1}%
\providecommand \bibfnamefont [1]{#1}%
\providecommand \citenamefont [1]{#1}%
\providecommand \href@noop [0]{\@secondoftwo}%
\providecommand \href [0]{\begingroup \@sanitize@url \@href}%
\providecommand \@href[1]{\@@startlink{#1}\@@href}%
\providecommand \@@href[1]{\endgroup#1\@@endlink}%
\providecommand \@sanitize@url [0]{\catcode `\\12\catcode `\$12\catcode
  `\&12\catcode `\#12\catcode `\^12\catcode `\_12\catcode `\%12\relax}%
\providecommand \@@startlink[1]{}%
\providecommand \@@endlink[0]{}%
\providecommand \url  [0]{\begingroup\@sanitize@url \@url }%
\providecommand \@url [1]{\endgroup\@href {#1}{\urlprefix }}%
\providecommand \urlprefix  [0]{URL }%
\providecommand \Eprint [0]{\href }%
\providecommand \doibase [0]{http://dx.doi.org/}%
\providecommand \selectlanguage [0]{\@gobble}%
\providecommand \bibinfo  [0]{\@secondoftwo}%
\providecommand \bibfield  [0]{\@secondoftwo}%
\providecommand \translation [1]{[#1]}%
\providecommand \BibitemOpen [0]{}%
\providecommand \bibitemStop [0]{}%
\providecommand \bibitemNoStop [0]{.\EOS\space}%
\providecommand \EOS [0]{\spacefactor3000\relax}%
\providecommand \BibitemShut  [1]{\csname bibitem#1\endcsname}%
\let\auto@bib@innerbib\@empty
%</preamble>
\bibitem [{\citenamefont {Cavagna}(2009)}]{cavagna_supercooled_2009}%
  \BibitemOpen
  \bibfield  {author} {\bibinfo {author} {\bibfnamefont {A.}~\bibnamefont
  {Cavagna}},\ }\href {\doibase 10.1016/j.physrep.2009.03.003} {\bibfield
  {journal} {\bibinfo  {journal} {Phys. Rep.}\ }\textbf {\bibinfo {volume}
  {476}},\ \bibinfo {pages} {51} (\bibinfo {year} {2009})}\BibitemShut
  {NoStop}%
\bibitem [{\citenamefont {Berthier}\ and\ \citenamefont
  {Biroli}(2011)}]{berthier_theoretical_2011}%
  \BibitemOpen
  \bibfield  {author} {\bibinfo {author} {\bibfnamefont {L.}~\bibnamefont
  {Berthier}}\ and\ \bibinfo {author} {\bibfnamefont {G.}~\bibnamefont
  {Biroli}},\ }\href {\doibase 10.1103/RevModPhys.83.587} {\bibfield  {journal}
  {\bibinfo  {journal} {Rev. Mod. Phys.}\ }\textbf {\bibinfo {volume} {83}},\
  \bibinfo {pages} {587} (\bibinfo {year} {2011})}\BibitemShut {NoStop}%
\bibitem [{\citenamefont {Gotze}(2009)}]{gotze_complex_2009}%
  \BibitemOpen
  \bibfield  {author} {\bibinfo {author} {\bibfnamefont {W.}~\bibnamefont
  {Gotze}},\ }\href@noop {} {\emph {\bibinfo {title} {Complex {Dynamics} of
  {Glass}-{Forming} {Liquids}: {A} {Mode}-{Coupling} {Theory}}}}\ (\bibinfo
  {publisher} {Oxford University Press, USA},\ \bibinfo {year}
  {2009})\BibitemShut {NoStop}%
\bibitem [{\citenamefont {Szamel}(2003)}]{Szamel_2003}%
  \BibitemOpen
  \bibfield  {author} {\bibinfo {author} {\bibfnamefont {G.}~\bibnamefont
  {Szamel}},\ }\href {\doibase 10.1103/PhysRevLett.90.228301} {\bibfield
  {journal} {\bibinfo  {journal} {Phys. Rev. Lett.}\ }\textbf {\bibinfo
  {volume} {90}},\ \bibinfo {pages} {228301} (\bibinfo {year}
  {2003})}\BibitemShut {NoStop}%
\bibitem [{\citenamefont {Mayer}\ \emph {et~al.}(2006)\citenamefont {Mayer},
  \citenamefont {Miyazaki},\ and\ \citenamefont {Reichman}}]{Mayer2006}%
  \BibitemOpen
  \bibfield  {author} {\bibinfo {author} {\bibfnamefont {P.}~\bibnamefont
  {Mayer}}, \bibinfo {author} {\bibfnamefont {K.}~\bibnamefont {Miyazaki}}, \
  and\ \bibinfo {author} {\bibfnamefont {D.~R.}\ \bibnamefont {Reichman}},\
  }\href {\doibase 10.1103/PhysRevLett.97.095702} {\bibfield  {journal}
  {\bibinfo  {journal} {Phys. Rev. Lett.}\ }\textbf {\bibinfo {volume} {97}},\
  \bibinfo {pages} {095702} (\bibinfo {year} {2006})}\BibitemShut {NoStop}%
\bibitem [{\citenamefont {Janssen}\ and\ \citenamefont
  {Reichman}(2015)}]{janssen_microscopic_2015}%
  \BibitemOpen
  \bibfield  {author} {\bibinfo {author} {\bibfnamefont {L.~M.}\ \bibnamefont
  {Janssen}}\ and\ \bibinfo {author} {\bibfnamefont {D.~R.}\ \bibnamefont
  {Reichman}},\ }\href {\doibase 10.1103/PhysRevLett.115.205701} {\bibfield
  {journal} {\bibinfo  {journal} {Phys. Rev. Lett.}\ }\textbf {\bibinfo
  {volume} {115}},\ \bibinfo {pages} {205701} (\bibinfo {year}
  {2015})}\BibitemShut {NoStop}%
\bibitem [{\citenamefont {Luo}\ \emph {et~al.}(2021)\citenamefont {Luo},
  \citenamefont {Debets},\ and\ \citenamefont
  {Janssen}}]{luo_tagged-particle_2021}%
  \BibitemOpen
  \bibfield  {author} {\bibinfo {author} {\bibfnamefont {C.}~\bibnamefont
  {Luo}}, \bibinfo {author} {\bibfnamefont {V.~E.}\ \bibnamefont {Debets}}, \
  and\ \bibinfo {author} {\bibfnamefont {L.~M.~C.}\ \bibnamefont {Janssen}},\
  }\href {\doibase 10.1063/5.0056257} {\bibfield  {journal} {\bibinfo
  {journal} {J. Chem. Phys.}\ }\textbf {\bibinfo {volume} {155}},\ \bibinfo
  {pages} {034502} (\bibinfo {year} {2021})}\BibitemShut {NoStop}%
\bibitem [{\citenamefont {Ciarella}\ \emph {et~al.}(2021)\citenamefont
  {Ciarella}, \citenamefont {Luo}, \citenamefont {Debets},\ and\ \citenamefont
  {Janssen}}]{ciarella_multi-component_2021}%
  \BibitemOpen
  \bibfield  {author} {\bibinfo {author} {\bibfnamefont {S.}~\bibnamefont
  {Ciarella}}, \bibinfo {author} {\bibfnamefont {C.}~\bibnamefont {Luo}},
  \bibinfo {author} {\bibfnamefont {V.~E.}\ \bibnamefont {Debets}}, \ and\
  \bibinfo {author} {\bibfnamefont {L.~M.~C.}\ \bibnamefont {Janssen}},\ }\href
  {\doibase 10.1140/epje/s10189-021-00095-w} {\bibfield  {journal} {\bibinfo
  {journal} {Eur. Phys. J. E}\ }\textbf {\bibinfo {volume} {44}},\ \bibinfo
  {pages} {91} (\bibinfo {year} {2021})}\BibitemShut {NoStop}%
\bibitem [{\citenamefont {Baity-Jesi}\ and\ \citenamefont
  {Reichman}(2019)}]{Baity-Jesi_Reichman_2019}%
  \BibitemOpen
  \bibfield  {author} {\bibinfo {author} {\bibfnamefont {M.}~\bibnamefont
  {Baity-Jesi}}\ and\ \bibinfo {author} {\bibfnamefont {D.~R.}\ \bibnamefont
  {Reichman}},\ }\href {\doibase 10.1063/1.5115042} {\bibfield  {journal}
  {\bibinfo  {journal} {J. Chem. Phys.}\ }\textbf {\bibinfo {volume} {151}},\
  \bibinfo {pages} {084503} (\bibinfo {year} {2019})}\BibitemShut {NoStop}%
\bibitem [{\citenamefont {Berthier}\ \emph {et~al.}(2020)\citenamefont
  {Berthier}, \citenamefont {Charbonneau},\ and\ \citenamefont
  {Kundu}}]{Berthier_Charbonneau_Kundu_2020}%
  \BibitemOpen
  \bibfield  {author} {\bibinfo {author} {\bibfnamefont {L.}~\bibnamefont
  {Berthier}}, \bibinfo {author} {\bibfnamefont {P.}~\bibnamefont
  {Charbonneau}}, \ and\ \bibinfo {author} {\bibfnamefont {J.}~\bibnamefont
  {Kundu}},\ }\href {\doibase 10.1103/PhysRevLett.125.108001} {\bibfield
  {journal} {\bibinfo  {journal} {Phys. Rev. Lett.}\ }\textbf {\bibinfo
  {volume} {125}},\ \bibinfo {pages} {108001} (\bibinfo {year}
  {2020})}\BibitemShut {NoStop}%
\bibitem [{\citenamefont {Maimbourg}\ \emph {et~al.}(2016)\citenamefont
  {Maimbourg}, \citenamefont {Kurchan},\ and\ \citenamefont
  {Zamponi}}]{Maimbourg_Kurchan_Zamponi_2016}%
  \BibitemOpen
  \bibfield  {author} {\bibinfo {author} {\bibfnamefont {T.}~\bibnamefont
  {Maimbourg}}, \bibinfo {author} {\bibfnamefont {J.}~\bibnamefont {Kurchan}},
  \ and\ \bibinfo {author} {\bibfnamefont {F.}~\bibnamefont {Zamponi}},\ }\href
  {\doibase 10.1103/PhysRevLett.116.015902} {\bibfield  {journal} {\bibinfo
  {journal} {Phys. Rev. Lett.}\ }\textbf {\bibinfo {volume} {116}},\ \bibinfo
  {pages} {015902} (\bibinfo {year} {2016})}\BibitemShut {NoStop}%
\bibitem [{\citenamefont {Stillinger}\ and\ \citenamefont
  {Weber}(1982)}]{Stillinger1982}%
  \BibitemOpen
  \bibfield  {author} {\bibinfo {author} {\bibfnamefont {F.~H.}\ \bibnamefont
  {Stillinger}}\ and\ \bibinfo {author} {\bibfnamefont {T.~A.}\ \bibnamefont
  {Weber}},\ }\href {\doibase 10.1103/PhysRevA.25.978} {\bibfield  {journal}
  {\bibinfo  {journal} {Phys. Rev. A}\ }\textbf {\bibinfo {volume} {25}},\
  \bibinfo {pages} {978} (\bibinfo {year} {1982})}\BibitemShut {NoStop}%
\bibitem [{\citenamefont {Wales}\ \emph {et~al.}(2003)\citenamefont {Wales},
  \citenamefont {Saykally}, \citenamefont {Zewail},\ and\ \citenamefont
  {King}}]{wales2003energy}%
  \BibitemOpen
  \bibfield  {author} {\bibinfo {author} {\bibfnamefont {D.}~\bibnamefont
  {Wales}}, \bibinfo {author} {\bibfnamefont {R.}~\bibnamefont {Saykally}},
  \bibinfo {author} {\bibfnamefont {A.}~\bibnamefont {Zewail}}, \ and\ \bibinfo
  {author} {\bibfnamefont {D.}~\bibnamefont {King}},\ }\href@noop {} {\emph
  {\bibinfo {title} {Energy Landscapes: Applications to Clusters, Biomolecules
  and Glasses}}}\ (\bibinfo  {publisher} {Cambridge University Press},\
  \bibinfo {year} {2003})\BibitemShut {NoStop}%
\bibitem [{\citenamefont {Sciortino}(2005)}]{Sciortino_2005}%
  \BibitemOpen
  \bibfield  {author} {\bibinfo {author} {\bibfnamefont {F.}~\bibnamefont
  {Sciortino}},\ }\href {\doibase 10.1088/1742-5468/2005/05/P05015} {\bibfield
  {journal} {\bibinfo  {journal} {J. Stat. Mech.}\ }\textbf {\bibinfo {volume}
  {2005}},\ \bibinfo {pages} {P05015} (\bibinfo {year} {2005})}\BibitemShut
  {NoStop}%
\bibitem [{\citenamefont {Heuer}(2008)}]{Heuer2008}%
  \BibitemOpen
  \bibfield  {author} {\bibinfo {author} {\bibfnamefont {A.}~\bibnamefont
  {Heuer}},\ }\href {\doibase 10.1088/0953-8984/20/37/373101} {\bibfield
  {journal} {\bibinfo  {journal} {J. Phys.: Condens. Matt.}\ }\textbf {\bibinfo
  {volume} {20}},\ \bibinfo {pages} {373101} (\bibinfo {year}
  {2008})}\BibitemShut {NoStop}%
\bibitem [{\citenamefont {Doye}\ and\ \citenamefont
  {Wales}(2002)}]{Doye_Saddle_2002}%
  \BibitemOpen
  \bibfield  {author} {\bibinfo {author} {\bibfnamefont {J.~P.~K.}\
  \bibnamefont {Doye}}\ and\ \bibinfo {author} {\bibfnamefont {D.~J.}\
  \bibnamefont {Wales}},\ }\href {\doibase 10.1063/1.1436470} {\bibfield
  {journal} {\bibinfo  {journal} {J. Chem. Phys.}\ }\textbf {\bibinfo {volume}
  {116}},\ \bibinfo {pages} {3777} (\bibinfo {year} {2002})}\BibitemShut
  {NoStop}%
\bibitem [{\citenamefont {Stratt}(1995)}]{Stratt_1995}%
  \BibitemOpen
  \bibfield  {author} {\bibinfo {author} {\bibfnamefont {R.~M.}\ \bibnamefont
  {Stratt}},\ }\href {\doibase 10.1021/ar00053a001} {\bibfield  {journal}
  {\bibinfo  {journal} {Acc. Chem. Res.}\ }\textbf {\bibinfo {volume} {28}},\
  \bibinfo {pages} {201} (\bibinfo {year} {1995})}\BibitemShut {NoStop}%
\bibitem [{\citenamefont {Bembenek}\ and\ \citenamefont
  {Laird}(1995)}]{Bembenek_Laird_1995}%
  \BibitemOpen
  \bibfield  {author} {\bibinfo {author} {\bibfnamefont {S.~D.}\ \bibnamefont
  {Bembenek}}\ and\ \bibinfo {author} {\bibfnamefont {B.~B.}\ \bibnamefont
  {Laird}},\ }\href {\doibase 10.1103/PhysRevLett.74.936} {\bibfield  {journal}
  {\bibinfo  {journal} {Phys. Rev. Lett.}\ }\textbf {\bibinfo {volume} {74}},\
  \bibinfo {pages} {936} (\bibinfo {year} {1995})}\BibitemShut {NoStop}%
\bibitem [{\citenamefont {Keyes}\ \emph {et~al.}(1997)\citenamefont {Keyes},
  \citenamefont {Vijayadamodar},\ and\ \citenamefont
  {Zurcher}}]{Keyes_Vijayadamodar_Zurcher_1997}%
  \BibitemOpen
  \bibfield  {author} {\bibinfo {author} {\bibfnamefont {T.}~\bibnamefont
  {Keyes}}, \bibinfo {author} {\bibfnamefont {G.~V.}\ \bibnamefont
  {Vijayadamodar}}, \ and\ \bibinfo {author} {\bibfnamefont {U.}~\bibnamefont
  {Zurcher}},\ }\href {\doibase doi:10.1063/1.473481} {\bibfield  {journal}
  {\bibinfo  {journal} {J. Chem. Phys.}\ }\textbf {\bibinfo {volume} {106}},\
  \bibinfo {pages} {4651} (\bibinfo {year} {1997})}\BibitemShut {NoStop}%
\bibitem [{\citenamefont {Ribeiro}\ and\ \citenamefont
  {Madden}(1997)}]{Ribeiro_Madden_1997}%
  \BibitemOpen
  \bibfield  {author} {\bibinfo {author} {\bibfnamefont {M.~C.~C.}\
  \bibnamefont {Ribeiro}}\ and\ \bibinfo {author} {\bibfnamefont {P.~A.}\
  \bibnamefont {Madden}},\ }\href {\doibase doi:10.1063/1.473917} {\bibfield
  {journal} {\bibinfo  {journal} {J. Chem. Phys.}\ }\textbf {\bibinfo {volume}
  {106}},\ \bibinfo {pages} {8616} (\bibinfo {year} {1997})}\BibitemShut
  {NoStop}%
\bibitem [{\citenamefont {Krämer}\ \emph {et~al.}(1998)\citenamefont
  {Krämer}, \citenamefont {Buchner},\ and\ \citenamefont
  {Dorfmüller}}]{Kramer_Buchner_Dorfmuller_1998}%
  \BibitemOpen
  \bibfield  {author} {\bibinfo {author} {\bibfnamefont {N.}~\bibnamefont
  {Krämer}}, \bibinfo {author} {\bibfnamefont {M.}~\bibnamefont {Buchner}}, \
  and\ \bibinfo {author} {\bibfnamefont {T.}~\bibnamefont {Dorfmüller}},\
  }\href {\doibase 10.1063/1.476768} {\bibfield  {journal} {\bibinfo  {journal}
  {J. Chem. Phys.}\ }\textbf {\bibinfo {volume} {109}},\ \bibinfo {pages}
  {1912} (\bibinfo {year} {1998})}\BibitemShut {NoStop}%
\bibitem [{\citenamefont {Gezelter}\ \emph {et~al.}(1997)\citenamefont
  {Gezelter}, \citenamefont {Rabani},\ and\ \citenamefont
  {Berne}}]{Gezelter_Rabani_Berne_1997}%
  \BibitemOpen
  \bibfield  {author} {\bibinfo {author} {\bibfnamefont {J.~D.}\ \bibnamefont
  {Gezelter}}, \bibinfo {author} {\bibfnamefont {E.}~\bibnamefont {Rabani}}, \
  and\ \bibinfo {author} {\bibfnamefont {B.~J.}\ \bibnamefont {Berne}},\ }\href
  {\doibase 10.1063/1.474822} {\bibfield  {journal} {\bibinfo  {journal} {J.
  Chem. Phys.}\ }\textbf {\bibinfo {volume} {107}},\ \bibinfo {pages} {4618}
  (\bibinfo {year} {1997})}\BibitemShut {NoStop}%
\bibitem [{\citenamefont {Donati}\ \emph {et~al.}(2000)\citenamefont {Donati},
  \citenamefont {Sciortino},\ and\ \citenamefont
  {Tartaglia}}]{Donati_Sciortino_Tartaglia_2000}%
  \BibitemOpen
  \bibfield  {author} {\bibinfo {author} {\bibfnamefont {C.}~\bibnamefont
  {Donati}}, \bibinfo {author} {\bibfnamefont {F.}~\bibnamefont {Sciortino}}, \
  and\ \bibinfo {author} {\bibfnamefont {P.}~\bibnamefont {Tartaglia}},\ }\href
  {\doibase 10.1103/PhysRevLett.85.1464} {\bibfield  {journal} {\bibinfo
  {journal} {Phys. Rev. Lett.}\ }\textbf {\bibinfo {volume} {85}},\ \bibinfo
  {pages} {1464} (\bibinfo {year} {2000})}\BibitemShut {NoStop}%
\bibitem [{\citenamefont {Clapa}\ \emph {et~al.}(2012)\citenamefont {Clapa},
  \citenamefont {Kottos},\ and\ \citenamefont
  {Starr}}]{Clapa_Kottos_Starr_2012}%
  \BibitemOpen
  \bibfield  {author} {\bibinfo {author} {\bibfnamefont {V.~I.}\ \bibnamefont
  {Clapa}}, \bibinfo {author} {\bibfnamefont {T.}~\bibnamefont {Kottos}}, \
  and\ \bibinfo {author} {\bibfnamefont {F.~W.}\ \bibnamefont {Starr}},\ }\href
  {\doibase doi:10.1063/1.3701564} {\bibfield  {journal} {\bibinfo  {journal}
  {J. Chem. Phys.}\ }\textbf {\bibinfo {volume} {136}},\ \bibinfo {pages}
  {144504} (\bibinfo {year} {2012})}\BibitemShut {NoStop}%
\bibitem [{\citenamefont {Zhang}\ \emph {et~al.}(2019)\citenamefont {Zhang},
  \citenamefont {Douglas},\ and\ \citenamefont
  {Starr}}]{Zhang_Douglas_Starr_2019}%
  \BibitemOpen
  \bibfield  {author} {\bibinfo {author} {\bibfnamefont {W.}~\bibnamefont
  {Zhang}}, \bibinfo {author} {\bibfnamefont {J.~F.}\ \bibnamefont {Douglas}},
  \ and\ \bibinfo {author} {\bibfnamefont {F.~W.}\ \bibnamefont {Starr}},\
  }\href {\doibase 10.1063/1.5127821} {\bibfield  {journal} {\bibinfo
  {journal} {J. Chem. Phys.}\ }\textbf {\bibinfo {volume} {151}},\ \bibinfo
  {pages} {184904} (\bibinfo {year} {2019})}\BibitemShut {NoStop}%
\bibitem [{\citenamefont {Kriuchevksyi}\ \emph {et~al.}(2021)\citenamefont
  {Kriuchevksyi}, \citenamefont {Sirk},\ and\ \citenamefont
  {Zaccone}}]{Kriuchevksyi_Sirk_Zaccone_2021}%
  \BibitemOpen
  \bibfield  {author} {\bibinfo {author} {\bibfnamefont {I.}~\bibnamefont
  {Kriuchevksyi}}, \bibinfo {author} {\bibfnamefont {T.}~\bibnamefont {Sirk}},
  \ and\ \bibinfo {author} {\bibfnamefont {A.}~\bibnamefont {Zaccone}},\ }\href
  {https://arxiv.org/abs/2105.00306} {\bibfield  {journal} {\bibinfo  {journal}
  {arXiv:2105.00306}\ } (\bibinfo {year} {2021})}\BibitemShut {NoStop}%
\bibitem [{\citenamefont {Cavagna}\ \emph {et~al.}(2003)\citenamefont
  {Cavagna}, \citenamefont {Giardina},\ and\ \citenamefont
  {Grigera}}]{cavagna_single_2003}%
  \BibitemOpen
  \bibfield  {author} {\bibinfo {author} {\bibfnamefont {A.}~\bibnamefont
  {Cavagna}}, \bibinfo {author} {\bibfnamefont {I.}~\bibnamefont {Giardina}}, \
  and\ \bibinfo {author} {\bibfnamefont {T.~S.}\ \bibnamefont {Grigera}},\
  }\href {\doibase 10.1088/0305-4470/36/43/004} {\bibfield  {journal} {\bibinfo
   {journal} {J. Phys. A: Math. Gen.}\ }\textbf {\bibinfo {volume} {36}},\
  \bibinfo {pages} {10721} (\bibinfo {year} {2003})}\BibitemShut {NoStop}%
\bibitem [{\citenamefont {Grigera}\ \emph {et~al.}(2002)\citenamefont
  {Grigera}, \citenamefont {Cavagna}, \citenamefont {Giardina},\ and\
  \citenamefont {Parisi}}]{grigera_geometric_2002}%
  \BibitemOpen
  \bibfield  {author} {\bibinfo {author} {\bibfnamefont {T.~S.}\ \bibnamefont
  {Grigera}}, \bibinfo {author} {\bibfnamefont {A.}~\bibnamefont {Cavagna}},
  \bibinfo {author} {\bibfnamefont {I.}~\bibnamefont {Giardina}}, \ and\
  \bibinfo {author} {\bibfnamefont {G.}~\bibnamefont {Parisi}},\ }\href
  {\doibase 10.1103/PhysRevLett.88.055502} {\bibfield  {journal} {\bibinfo
  {journal} {Phys. Rev. Lett.}\ }\textbf {\bibinfo {volume} {88}},\ \bibinfo
  {pages} {055502} (\bibinfo {year} {2002})}\BibitemShut {NoStop}%
\bibitem [{\citenamefont {Wales}\ and\ \citenamefont
  {Doye}(2001)}]{wales_dynamics_2001}%
  \BibitemOpen
  \bibfield  {author} {\bibinfo {author} {\bibfnamefont {D.~J.}\ \bibnamefont
  {Wales}}\ and\ \bibinfo {author} {\bibfnamefont {J.~P.~K.}\ \bibnamefont
  {Doye}},\ }\href {\doibase 10.1103/PhysRevB.63.214204} {\bibfield  {journal}
  {\bibinfo  {journal} {Phys. Rev. B}\ }\textbf {\bibinfo {volume} {63}},\
  \bibinfo {pages} {214204} (\bibinfo {year} {2001})}\BibitemShut {NoStop}%
\bibitem [{\citenamefont {Wales}\ and\ \citenamefont
  {Doye}(2003)}]{Wales_Doye_2003}%
  \BibitemOpen
  \bibfield  {author} {\bibinfo {author} {\bibfnamefont {D.~J.}\ \bibnamefont
  {Wales}}\ and\ \bibinfo {author} {\bibfnamefont {J.~P.~K.}\ \bibnamefont
  {Doye}},\ }\href {\doibase 10.1063/1.1625644} {\bibfield  {journal} {\bibinfo
   {journal} {J. Chem. Phys.}\ }\textbf {\bibinfo {volume} {119}},\ \bibinfo
  {pages} {12409} (\bibinfo {year} {2003})}\BibitemShut {NoStop}%
\bibitem [{\citenamefont {Grigera}(2006)}]{Grigera_2006}%
  \BibitemOpen
  \bibfield  {author} {\bibinfo {author} {\bibfnamefont {T.~S.}\ \bibnamefont
  {Grigera}},\ }\href {\doibase 10.1063/1.2151899} {\bibfield  {journal}
  {\bibinfo  {journal} {J. Chem. Phys.}\ }\textbf {\bibinfo {volume} {124}},\
  \bibinfo {pages} {064502} (\bibinfo {year} {2006})}\BibitemShut {NoStop}%
\bibitem [{\citenamefont {Coslovich}\ \emph {et~al.}(2019)\citenamefont
  {Coslovich}, \citenamefont {Ninarello},\ and\ \citenamefont
  {Berthier}}]{coslovich_localization_2019}%
  \BibitemOpen
  \bibfield  {author} {\bibinfo {author} {\bibfnamefont {D.}~\bibnamefont
  {Coslovich}}, \bibinfo {author} {\bibfnamefont {A.}~\bibnamefont
  {Ninarello}}, \ and\ \bibinfo {author} {\bibfnamefont {L.}~\bibnamefont
  {Berthier}},\ }\href {\doibase 10.21468/SciPostPhys.7.6.077} {\bibfield
  {journal} {\bibinfo  {journal} {SciPost Phys.}\ }\textbf {\bibinfo {volume}
  {7}},\ \bibinfo {pages} {077} (\bibinfo {year} {2019})}\BibitemShut {NoStop}%
\bibitem [{\citenamefont {Shimada}\ \emph {et~al.}(2021)\citenamefont
  {Shimada}, \citenamefont {Coslovich}, \citenamefont {Mizuno},\ and\
  \citenamefont {Ikeda}}]{Shimada_Coslovich_Mizuno_Ikeda_2021}%
  \BibitemOpen
  \bibfield  {author} {\bibinfo {author} {\bibfnamefont {M.}~\bibnamefont
  {Shimada}}, \bibinfo {author} {\bibfnamefont {D.}~\bibnamefont {Coslovich}},
  \bibinfo {author} {\bibfnamefont {H.}~\bibnamefont {Mizuno}}, \ and\ \bibinfo
  {author} {\bibfnamefont {A.}~\bibnamefont {Ikeda}},\ }\href {\doibase
  10.21468/SciPostPhys.10.1.001} {\bibfield  {journal} {\bibinfo  {journal}
  {SciPost Phys.}\ }\textbf {\bibinfo {volume} {10}},\ \bibinfo {pages} {001}
  (\bibinfo {year} {2021})}\BibitemShut {NoStop}%
\bibitem [{\citenamefont {Biroli}\ \emph
  {et~al.}(2006{\natexlab{a}})\citenamefont {Biroli}, \citenamefont {Bouchaud},
  \citenamefont {Miyazaki},\ and\ \citenamefont
  {Reichman}}]{Biroli_Bouchaud_Miyazaki_Reichman_2006}%
  \BibitemOpen
  \bibfield  {author} {\bibinfo {author} {\bibfnamefont {G.}~\bibnamefont
  {Biroli}}, \bibinfo {author} {\bibfnamefont {J.-P.}\ \bibnamefont
  {Bouchaud}}, \bibinfo {author} {\bibfnamefont {K.}~\bibnamefont {Miyazaki}},
  \ and\ \bibinfo {author} {\bibfnamefont {D.~R.}\ \bibnamefont {Reichman}},\
  }\href {\doibase 10.1103/PhysRevLett.97.195701} {\bibfield  {journal}
  {\bibinfo  {journal} {Phys. Rev. Lett.}\ }\textbf {\bibinfo {volume} {97}},\
  \bibinfo {pages} {195701} (\bibinfo {year} {2006}{\natexlab{a}})}\BibitemShut
  {NoStop}%
\bibitem [{\citenamefont {Zwanzig}(2001)}]{zwanzig2001nonequilibrium}%
  \BibitemOpen
  \bibfield  {author} {\bibinfo {author} {\bibfnamefont {R.}~\bibnamefont
  {Zwanzig}},\ }\href@noop {} {\emph {\bibinfo {title} {Nonequilibrium
  Statistical Mechanics}}}\ (\bibinfo  {publisher} {Oxford University Press},\
  \bibinfo {year} {2001})\BibitemShut {NoStop}%
\bibitem [{\citenamefont {Ikeda}\ \emph {et~al.}(2013)\citenamefont {Ikeda},
  \citenamefont {Berthier},\ and\ \citenamefont {Biroli}}]{Ikeda2013}%
  \BibitemOpen
  \bibfield  {author} {\bibinfo {author} {\bibfnamefont {A.}~\bibnamefont
  {Ikeda}}, \bibinfo {author} {\bibfnamefont {L.}~\bibnamefont {Berthier}}, \
  and\ \bibinfo {author} {\bibfnamefont {G.}~\bibnamefont {Biroli}},\ }\href
  {\doibase 10.1063/1.4769251} {\bibfield  {journal} {\bibinfo  {journal} {J.
  Chem. Phys.}\ }\textbf {\bibinfo {volume} {138}},\ \bibinfo {eid} {12A507}
  (\bibinfo {year} {2013})}\BibitemShut {NoStop}%
\bibitem [{\citenamefont {Widmer-Cooper}\ \emph {et~al.}(2004)\citenamefont
  {Widmer-Cooper}, \citenamefont {Harrowell},\ and\ \citenamefont
  {Fynewever}}]{Widmer-Cooper_Harrowell_Fynewever_2004}%
  \BibitemOpen
  \bibfield  {author} {\bibinfo {author} {\bibfnamefont {A.}~\bibnamefont
  {Widmer-Cooper}}, \bibinfo {author} {\bibfnamefont {P.}~\bibnamefont
  {Harrowell}}, \ and\ \bibinfo {author} {\bibfnamefont {H.}~\bibnamefont
  {Fynewever}},\ }\href {\doibase 10.1103/PhysRevLett.93.135701} {\bibfield
  {journal} {\bibinfo  {journal} {Phys. Rev. Lett.}\ }\textbf {\bibinfo
  {volume} {93}},\ \bibinfo {pages} {135701} (\bibinfo {year}
  {2004})}\BibitemShut {NoStop}%
\bibitem [{\citenamefont {Berthier}\ and\ \citenamefont
  {Jack}(2007)}]{Berthier_Jack_2007}%
  \BibitemOpen
  \bibfield  {author} {\bibinfo {author} {\bibfnamefont {L.}~\bibnamefont
  {Berthier}}\ and\ \bibinfo {author} {\bibfnamefont {R.~L.}\ \bibnamefont
  {Jack}},\ }\href {\doibase 10.1103/PhysRevE.76.041509} {\bibfield  {journal}
  {\bibinfo  {journal} {Phys. Rev. E}\ }\textbf {\bibinfo {volume} {76}},\
  \bibinfo {pages} {041509} (\bibinfo {year} {2007})}\BibitemShut {NoStop}%
\bibitem [{\citenamefont {Franz}\ \emph {et~al.}(2011)\citenamefont {Franz},
  \citenamefont {Parisi}, \citenamefont {Ricci-Tersenghi},\ and\ \citenamefont
  {Rizzo}}]{Franz2011a}%
  \BibitemOpen
  \bibfield  {author} {\bibinfo {author} {\bibfnamefont {S.}~\bibnamefont
  {Franz}}, \bibinfo {author} {\bibfnamefont {G.}~\bibnamefont {Parisi}},
  \bibinfo {author} {\bibfnamefont {F.}~\bibnamefont {Ricci-Tersenghi}}, \ and\
  \bibinfo {author} {\bibfnamefont {T.}~\bibnamefont {Rizzo}},\ }\href
  {https://doi.org/10.1140/epje/i2011-11102-0} {\bibfield  {journal} {\bibinfo
  {journal} {The European Physical Journal E}\ }\textbf {\bibinfo {volume}
  {34}},\ \bibinfo {pages} {102} (\bibinfo {year} {2011})}\BibitemShut
  {NoStop}%
\bibitem [{\citenamefont {Mehta}(2004)}]{mehta2004random}%
  \BibitemOpen
  \bibfield  {author} {\bibinfo {author} {\bibfnamefont {M.}~\bibnamefont
  {Mehta}},\ }\href@noop {} {\emph {\bibinfo {title} {Random Matrices}}},\
  ISSN\ (\bibinfo  {publisher} {Elsevier Science},\ \bibinfo {year}
  {2004})\BibitemShut {NoStop}%
\bibitem [{\citenamefont {Cavagna}\ \emph {et~al.}(1998)\citenamefont
  {Cavagna}, \citenamefont {Giardina},\ and\ \citenamefont
  {Parisi}}]{Cavagna1998}%
  \BibitemOpen
  \bibfield  {author} {\bibinfo {author} {\bibfnamefont {A.}~\bibnamefont
  {Cavagna}}, \bibinfo {author} {\bibfnamefont {I.}~\bibnamefont {Giardina}}, \
  and\ \bibinfo {author} {\bibfnamefont {G.}~\bibnamefont {Parisi}},\ }\href
  {\doibase 10.1103/PhysRevB.57.11251} {\bibfield  {journal} {\bibinfo
  {journal} {Phys. Rev. B}\ }\textbf {\bibinfo {volume} {57}},\ \bibinfo
  {pages} {11251} (\bibinfo {year} {1998})}\BibitemShut {NoStop}%
\bibitem [{\citenamefont {Hansen}\ and\ \citenamefont
  {McDonald}(2006)}]{hansen_theory_2006}%
  \BibitemOpen
  \bibfield  {author} {\bibinfo {author} {\bibfnamefont {J.-P.}\ \bibnamefont
  {Hansen}}\ and\ \bibinfo {author} {\bibfnamefont {I.~R.}\ \bibnamefont
  {McDonald}},\ }\href@noop {} {\emph {\bibinfo {title} {Theory of {Simple}
  {Liquids}}}},\ \bibinfo {edition} {3rd}\ ed.\ (\bibinfo  {publisher}
  {Academic Press},\ \bibinfo {year} {2006})\BibitemShut {NoStop}%
\bibitem [{\citenamefont {Vineyard}(1958)}]{Vineyard_1958}%
  \BibitemOpen
  \bibfield  {author} {\bibinfo {author} {\bibfnamefont {G.~H.}\ \bibnamefont
  {Vineyard}},\ }\href {\doibase 10.1103/PhysRev.110.999} {\bibfield  {journal}
  {\bibinfo  {journal} {Phys. Rev.}\ }\textbf {\bibinfo {volume} {110}},\
  \bibinfo {pages} {999} (\bibinfo {year} {1958})}\BibitemShut {NoStop}%
\bibitem [{\citenamefont {Biroli}\ \emph
  {et~al.}(2006{\natexlab{b}})\citenamefont {Biroli}, \citenamefont {Bouchaud},
  \citenamefont {Miyazaki},\ and\ \citenamefont {Reichman}}]{Biroli2006}%
  \BibitemOpen
  \bibfield  {author} {\bibinfo {author} {\bibfnamefont {G.}~\bibnamefont
  {Biroli}}, \bibinfo {author} {\bibfnamefont {J.-P.}\ \bibnamefont
  {Bouchaud}}, \bibinfo {author} {\bibfnamefont {K.}~\bibnamefont {Miyazaki}},
  \ and\ \bibinfo {author} {\bibfnamefont {D.~R.}\ \bibnamefont {Reichman}},\
  }\href {\doibase 10.1103/PhysRevLett.97.195701} {\bibfield  {journal}
  {\bibinfo  {journal} {Phys. Rev. Lett.}\ }\textbf {\bibinfo {volume} {97}},\
  \bibinfo {pages} {195701} (\bibinfo {year} {2006}{\natexlab{b}})}\BibitemShut
  {NoStop}%
\bibitem [{\citenamefont {Berthier}\ \emph {et~al.}(2007)\citenamefont
  {Berthier}, \citenamefont {Biroli}, \citenamefont {Bouchaud}, \citenamefont
  {Kob}, \citenamefont {Miyazaki},\ and\ \citenamefont
  {Reichman}}]{Berthier2007a}%
  \BibitemOpen
  \bibfield  {author} {\bibinfo {author} {\bibfnamefont {L.}~\bibnamefont
  {Berthier}}, \bibinfo {author} {\bibfnamefont {G.}~\bibnamefont {Biroli}},
  \bibinfo {author} {\bibfnamefont {J.-P.}\ \bibnamefont {Bouchaud}}, \bibinfo
  {author} {\bibfnamefont {W.}~\bibnamefont {Kob}}, \bibinfo {author}
  {\bibfnamefont {K.}~\bibnamefont {Miyazaki}}, \ and\ \bibinfo {author}
  {\bibfnamefont {D.~R.}\ \bibnamefont {Reichman}},\ }\href {\doibase
  10.1063/1.2721554} {\bibfield  {journal} {\bibinfo  {journal} {J. Chem.
  Phys.}\ }\textbf {\bibinfo {volume} {126}},\ \bibinfo {eid} {184503}
  (\bibinfo {year} {2007})}\BibitemShut {NoStop}%
\bibitem [{\citenamefont {Rizzo}\ and\ \citenamefont
  {Voigtmann}(2020)}]{Rizzo_Voigtmann_2020}%
  \BibitemOpen
  \bibfield  {author} {\bibinfo {author} {\bibfnamefont {T.}~\bibnamefont
  {Rizzo}}\ and\ \bibinfo {author} {\bibfnamefont {T.}~\bibnamefont
  {Voigtmann}},\ }\href {\doibase 10.1103/PhysRevLett.124.195501} {\bibfield
  {journal} {\bibinfo  {journal} {Phys. Rev. Lett.}\ }\textbf {\bibinfo
  {volume} {124}},\ \bibinfo {pages} {195501} (\bibinfo {year}
  {2020})}\BibitemShut {NoStop}%
\bibitem [{\citenamefont {Crisanti}\ \emph {et~al.}(1993)\citenamefont
  {Crisanti}, \citenamefont {Horner},\ and\ \citenamefont
  {Sommers}}]{Crisanti1993}%
  \BibitemOpen
  \bibfield  {author} {\bibinfo {author} {\bibfnamefont {A.}~\bibnamefont
  {Crisanti}}, \bibinfo {author} {\bibfnamefont {H.}~\bibnamefont {Horner}}, \
  and\ \bibinfo {author} {\bibfnamefont {H.~J.}\ \bibnamefont {Sommers}},\
  }\href@noop {} {\bibfield  {journal} {\bibinfo  {journal} {Z. Phys. B}\
  }\textbf {\bibinfo {volume} {92}},\ \bibinfo {pages} {257} (\bibinfo {year}
  {1993})}\BibitemShut {NoStop}%
\bibitem [{\citenamefont {Coslovich}\ and\ \citenamefont
  {Ikeda}(2021)}]{zenodo}%
  \BibitemOpen
  \bibfield  {author} {\bibinfo {author} {\bibfnamefont {D.}~\bibnamefont
  {Coslovich}}\ and\ \bibinfo {author} {\bibfnamefont {A.}~\bibnamefont
  {Ikeda}},\ }\href@noop {} {\enquote {\bibinfo {title} {Revisiting the
  single-saddle model for the $\beta$-relaxation of supercooled liquids},}\
  }\bibinfo {howpublished} {\url{https://doi.org/10.5281/zenodo.5791675}}
  (\bibinfo {year} {2021})\BibitemShut {NoStop}%
\bibitem [{\citenamefont {Gutiérrez}\ \emph {et~al.}(2015)\citenamefont
  {Gutiérrez}, \citenamefont {Karmakar}, \citenamefont {Pollack},\ and\
  \citenamefont {Procaccia}}]{gutierrez_static_2015}%
  \BibitemOpen
  \bibfield  {author} {\bibinfo {author} {\bibfnamefont {R.}~\bibnamefont
  {Gutiérrez}}, \bibinfo {author} {\bibfnamefont {S.}~\bibnamefont
  {Karmakar}}, \bibinfo {author} {\bibfnamefont {Y.~G.}\ \bibnamefont
  {Pollack}}, \ and\ \bibinfo {author} {\bibfnamefont {I.}~\bibnamefont
  {Procaccia}},\ }\href {\doibase 10.1209/0295-5075/111/56009} {\bibfield
  {journal} {\bibinfo  {journal} {EPL}\ }\textbf {\bibinfo {volume} {111}},\
  \bibinfo {pages} {56009} (\bibinfo {year} {2015})}\BibitemShut {NoStop}%
\bibitem [{\citenamefont {Ninarello}\ \emph {et~al.}(2017)\citenamefont
  {Ninarello}, \citenamefont {Berthier},\ and\ \citenamefont
  {Coslovich}}]{ninarello_models_2017}%
  \BibitemOpen
  \bibfield  {author} {\bibinfo {author} {\bibfnamefont {A.}~\bibnamefont
  {Ninarello}}, \bibinfo {author} {\bibfnamefont {L.}~\bibnamefont {Berthier}},
  \ and\ \bibinfo {author} {\bibfnamefont {D.}~\bibnamefont {Coslovich}},\
  }\href {\doibase 10.1103/PhysRevX.7.021039} {\bibfield  {journal} {\bibinfo
  {journal} {Phys. Rev. X}\ }\textbf {\bibinfo {volume} {7}},\ \bibinfo {pages}
  {021039} (\bibinfo {year} {2017})}\BibitemShut {NoStop}%
\bibitem [{\citenamefont {Gleim}\ \emph {et~al.}(1998)\citenamefont {Gleim},
  \citenamefont {Kob},\ and\ \citenamefont {Binder}}]{Gleim1998}%
  \BibitemOpen
  \bibfield  {author} {\bibinfo {author} {\bibfnamefont {T.}~\bibnamefont
  {Gleim}}, \bibinfo {author} {\bibfnamefont {W.}~\bibnamefont {Kob}}, \ and\
  \bibinfo {author} {\bibfnamefont {K.}~\bibnamefont {Binder}},\ }\href
  {\doibase 10.1103/PhysRevLett.81.4404} {\bibfield  {journal} {\bibinfo
  {journal} {Phys. Rev. Lett.}\ }\textbf {\bibinfo {volume} {81}},\ \bibinfo
  {pages} {4404} (\bibinfo {year} {1998})}\BibitemShut {NoStop}%
\bibitem [{\citenamefont {Kob}\ \emph {et~al.}(2002)\citenamefont {Kob},
  \citenamefont {Nauroth},\ and\ \citenamefont
  {Sciortino}}]{Kob_Nauroth_Sciortino_2002}%
  \BibitemOpen
  \bibfield  {author} {\bibinfo {author} {\bibfnamefont {W.}~\bibnamefont
  {Kob}}, \bibinfo {author} {\bibfnamefont {M.}~\bibnamefont {Nauroth}}, \ and\
  \bibinfo {author} {\bibfnamefont {F.}~\bibnamefont {Sciortino}},\ }\href
  {\doibase 10.1016/S0022-3093(02)01457} {\bibfield  {journal} {\bibinfo
  {journal} {J. Non-Cryst. Solids}\ }\textbf {\bibinfo {volume} {307}},\
  \bibinfo {pages} {181} (\bibinfo {year} {2002})}\BibitemShut {NoStop}%
\bibitem [{\citenamefont {Hocky}\ \emph {et~al.}(2014)\citenamefont {Hocky},
  \citenamefont {Coslovich}, \citenamefont {Ikeda},\ and\ \citenamefont
  {Reichman}}]{hocky_correlation_2014}%
  \BibitemOpen
  \bibfield  {author} {\bibinfo {author} {\bibfnamefont {G.~M.}\ \bibnamefont
  {Hocky}}, \bibinfo {author} {\bibfnamefont {D.}~\bibnamefont {Coslovich}},
  \bibinfo {author} {\bibfnamefont {A.}~\bibnamefont {Ikeda}}, \ and\ \bibinfo
  {author} {\bibfnamefont {D.~R.}\ \bibnamefont {Reichman}},\ }\href {\doibase
  10.1103/PhysRevLett.113.157801} {\bibfield  {journal} {\bibinfo  {journal}
  {Phys. Rev. Lett.}\ }\textbf {\bibinfo {volume} {113}},\ \bibinfo {pages}
  {157801} (\bibinfo {year} {2014})}\BibitemShut {NoStop}%
\bibitem [{\citenamefont {Paret}\ \emph {et~al.}(2020)\citenamefont {Paret},
  \citenamefont {Jack},\ and\ \citenamefont
  {Coslovich}}]{Paret_Jack_Coslovich_2020}%
  \BibitemOpen
  \bibfield  {author} {\bibinfo {author} {\bibfnamefont {J.}~\bibnamefont
  {Paret}}, \bibinfo {author} {\bibfnamefont {R.~L.}\ \bibnamefont {Jack}}, \
  and\ \bibinfo {author} {\bibfnamefont {D.}~\bibnamefont {Coslovich}},\ }\href
  {\doibase 10.1063/5.0004732} {\bibfield  {journal} {\bibinfo  {journal} {J.
  Chem. Phys.}\ }\textbf {\bibinfo {volume} {152}},\ \bibinfo {pages} {144502}
  (\bibinfo {year} {2020})}\BibitemShut {NoStop}%
\bibitem [{\citenamefont {Boattini}\ \emph {et~al.}(2020)\citenamefont
  {Boattini}, \citenamefont {Marín-Aguilar}, \citenamefont {Mitra},
  \citenamefont {Foffi}, \citenamefont {Smallenburg},\ and\ \citenamefont
  {Filion}}]{Boattini_Marin-Aguilar_Mitra_Foffi_Smallenburg_Filion_2020}%
  \BibitemOpen
  \bibfield  {author} {\bibinfo {author} {\bibfnamefont {E.}~\bibnamefont
  {Boattini}}, \bibinfo {author} {\bibfnamefont {S.}~\bibnamefont
  {Marín-Aguilar}}, \bibinfo {author} {\bibfnamefont {S.}~\bibnamefont
  {Mitra}}, \bibinfo {author} {\bibfnamefont {G.}~\bibnamefont {Foffi}},
  \bibinfo {author} {\bibfnamefont {F.}~\bibnamefont {Smallenburg}}, \ and\
  \bibinfo {author} {\bibfnamefont {L.}~\bibnamefont {Filion}},\ }\href
  {\doibase 10.1038/s41467-020-19286} {\bibfield  {journal} {\bibinfo
  {journal} {Nat. Comm.}\ }\textbf {\bibinfo {volume} {11}},\ \bibinfo {pages}
  {5479} (\bibinfo {year} {2020})}\BibitemShut {NoStop}%
\bibitem [{\citenamefont {Boattini}\ \emph {et~al.}(2021)\citenamefont
  {Boattini}, \citenamefont {Smallenburg},\ and\ \citenamefont
  {Filion}}]{Boattini_Smallenburg_Filion_2021}%
  \BibitemOpen
  \bibfield  {author} {\bibinfo {author} {\bibfnamefont {E.}~\bibnamefont
  {Boattini}}, \bibinfo {author} {\bibfnamefont {F.}~\bibnamefont
  {Smallenburg}}, \ and\ \bibinfo {author} {\bibfnamefont {L.}~\bibnamefont
  {Filion}},\ }\href {\doibase 10.1103/PhysRevLett.127.088007} {\bibfield
  {journal} {\bibinfo  {journal} {Phys. Rev. Lett.}\ }\textbf {\bibinfo
  {volume} {127}},\ \bibinfo {pages} {088007} (\bibinfo {year}
  {2021})}\BibitemShut {NoStop}%
\bibitem [{\citenamefont {Jack}\ \emph {et~al.}(2014)\citenamefont {Jack},
  \citenamefont {Dunleavy},\ and\ \citenamefont
  {Royall}}]{jack_information-theoretic_2014}%
  \BibitemOpen
  \bibfield  {author} {\bibinfo {author} {\bibfnamefont {R.~L.}\ \bibnamefont
  {Jack}}, \bibinfo {author} {\bibfnamefont {A.~J.}\ \bibnamefont {Dunleavy}},
  \ and\ \bibinfo {author} {\bibfnamefont {C.~P.}\ \bibnamefont {Royall}},\
  }\href {\doibase 10.1103/PhysRevLett.113.095703} {\bibfield  {journal}
  {\bibinfo  {journal} {Phys. Rev. Lett.}\ }\textbf {\bibinfo {volume} {113}},\
  \bibinfo {pages} {095703} (\bibinfo {year} {2014})}\BibitemShut {NoStop}%
\bibitem [{\citenamefont {Rizzo}(2016)}]{Rizzo_2016}%
  \BibitemOpen
  \bibfield  {author} {\bibinfo {author} {\bibfnamefont {T.}~\bibnamefont
  {Rizzo}},\ }\href {\doibase 10.1103/PhysRevB.94.014202} {\bibfield  {journal}
  {\bibinfo  {journal} {Phys. Rev. B}\ }\textbf {\bibinfo {volume} {94}},\
  \bibinfo {pages} {014202} (\bibinfo {year} {2016})}\BibitemShut {NoStop}%
\end{thebibliography}%
\bibliographystyle{apsrev4-1}

\end{document}